\documentclass[11pt,a4paper]{article}
\usepackage[utf8]{inputenc}
\usepackage[english]{babel}
\usepackage{amsmath}
\usepackage{amsfonts}
\usepackage{amssymb}
\usepackage{graphicx}
\usepackage[left=2.0cm,right=2.0cm,top=2.0cm,bottom=2.0cm]{geometry}
\usepackage{authblk} 
\usepackage[colorlinks=true, citecolor=blue, linkcolor=black, urlcolor=blue, filecolor=blue]{hyperref}
\usepackage{natbib}
\usepackage[section]{placeins}
\usepackage{breakcites}
\usepackage{array}
\usepackage{multirow}
\usepackage[T1]{fontenc}
\usepackage{xcolor}
\usepackage{enumitem}
\usepackage[font=small,skip=0pt]{caption}
\usepackage{subfigure}
\usepackage{soul}
\usepackage{float}
\usepackage{booktabs}
\usepackage[capitalise,nameinlink]{cleveref}
\crefname{figure}{Figure}{Figures}
\crefname{table}{Table}{Tables}
\usepackage{multirow}
\usepackage{siunitx}    
\usepackage{threeparttable}

\hypersetup{
    colorlinks=true,
    linkcolor=blue,         
    citecolor=blue,         
    urlcolor=blue           
}

\makeatother
\usepackage{sectsty}
\sectionfont{\large}
\subsectionfont{\large\normalfont\itshape}
\subsubsectionfont{\normalfont\itshape}

\title{Thermo-Hydro-Mechanical Modeling of a TBM Tunnel in Opalinus Clay: Thermal Coupling Effects on Long-term Lining Pressures and Excavation Damage Zone Development}

\author{Saeed Tourchi$^{1}$,
        Martin Ziegler$^{2,3}$,
        and Arash Alimardani Lavasan$^{1}$\thanks{Corresponding author: \texttt{arash.lavasan@uni.lu}}}

\affil{$^{1}$ Department of Engineering, University of Luxembourg, Luxembourg}
\affil{$^{2}$ Department of Earth and Planetary Sciences, ETH Zurich, Zurich, Switzerland}
\affil{$^{3}$ Federal Office of Topography (swisstopo), St. Ursanne, Switzerland}

\date{}

\begin{document}

\maketitle

\begin{abstract}
\noindent Tunnels in argillaceous formations such as Opalinus Clay (OPA) develop a coupled thermal, hydraulic, and mechanical response. This study examines the New Belchen Tunnel (STB), excavated by tunnel boring machine through faulted OPA shale. Field measurements show a strong relation between lining temperature and radial pressure. We use a two-dimensional plane-strain THM model to examine excavation, construction heat, seasonal temperature cycles, pore-pressure transients, and stress redistribution. The model reproduces the phasing and approximate magnitude of the seasonal pressure cycles and shows how the thermal expansion mismatch between the lining, grout, and OPA modifies lining load and near-field pore pressure. It underestimates the monotonic long-term pressure increase at several sensors. Comparison with radial extensometer measurements indicates that moisture-driven swelling, local geological structure, grout or interface damage, and three-dimensional construction effects also contribute. The model therefore supports a thermal interpretation of the cyclic response while defining the processes required for a fuller long-term back-analysis.
\\

\end{abstract}

\section{Introduction} \label{sec_introduction}

The long-term performance and safety of underground structures in argillaceous formations, such as tunnels, deep geological repositories, and underground storage facilities, depend critically on understanding the coupled thermo-hydro-mechanical (THM) behavior of the rock mass and its interaction with the lining \citep{Yu2010,tourchi2020thm,bumbieler2021,Golchin_2022a,manica2022a,Zhou2023,song2025thm}. Among the most studied formations in this regard is Opalinus Clay (OPA), a Jurassic-aged claystone (mudstone) that has attracted extensive interest due to its low hydraulic conductivity, self-sealing capacity, radionuclide retention potential, and other favorable properties for long-term nuclear waste storage; mechanically, the material can be challenging. These attributes have led to its selection as the reference host rock in Switzerland for the deep geological disposal of high-level radioactive waste \citep{gens2007,Tsang2012,Nagra_NTB_24-17_2024}. The OPA formation is also relevant for tunnel infrastructure projects.

However, excavation in such materials, particularly when using Tunnel Boring Machines (TBMs), disturbs the pre-existing stress equilibrium, initiating complex time-dependent processes in the surrounding rock mass \citep{li2013numerical,song2021hydro}. These include (i) mechanical unloading and damage leading to stress redistribution, (ii) hydraulic gradients due to pore-pressure changes, desaturation, and resaturation processes, and (iii) thermal perturbations arising from construction processes, lining hydration, or environmental interaction. Consequently, an excavation damage zone (EDZ) evolves and is influenced by coupled interactions among these processes, often manifesting as delayed deformation, stress recovery, and volumetric changes, including swelling \citep{bossart_geological_2002,kwon2008influence,Gens2013,bossart2017,tourchi2019coupled,tourchi2019thermo}. Early analyses of tunnel stability emphasized mechanical and hydromechanical interactions \citep{einstein2000,franzius_turning_2011,lavasan2018} under the assumption that the thermal effects were negligible at the time. However, as infrastructure is pushed deeper underground and analogies from nuclear repository research are increasingly adopted in civil engineering, it is evident that transient thermal effects play a critical role in stress redistribution and material behavior \citep{Francois2009,Gens2010,tourchi_thermomechanical_2023}.

In the case of Opalinus Clay, laboratory and in situ experimental programs, such as those at the Mont Terri Underground Rock Laboratory (MT URL), have demonstrated that the material exhibits a strong coupling between temperature, pore pressure, and mechanical behavior \citep{bossart2008,bossart2017,ziegler2026mttm43}. Key features include extremely low permeability, anisotropic stiffness and strength owing to bedding orientation, time-dependent swelling, and thermally induced pore-pressure changes under undrained conditions \citep{Hueckel1990,Armand2004,naumann_experimental_2007,Monfared2014,Favero_2016,wild_experimental_2018,song2025constitutive}. These characteristics lead to behaviors that simplified conventional methods cannot accurately capture. Recent advances in constitutive modeling of argillaceous rocks have demonstrated the importance of incorporating thermal coupling and anisotropic behavior in numerical simulations \citep{manica2022part1,manica2022part2,tourchi2023thermo}.

Despite this growing body of knowledge, real-world case studies combining long-term field monitoring with high-fidelity THM simulations remain rare, especially in TBM-excavated tunnels, where excavation is continuous and multiphysical interactions evolve concurrently. A particularly relevant and well-documented case is the New Belchen Tunnel (STB; \textit{Sanierungstunnel Belchen}) in the Swiss Jura Mountains, excavated using TBM technology and instrumented within a comprehensive monitoring program \citep{Grob1972,Amstad2001,ziegler2017,ziegler2018,renz2019ensi,ziegler2020ensi,ziegler2021ensi,ZIEGLER-Arash2022,ziegler2022ensi,ziegler2023final}. The STB intersects the Opalinus Clay at a maximum overburden of approximately 325\,m. The radial pressures measured at the STB lining exhibited delayed increases post-excavation, suggestive of combined swelling and thermal effects. However, decoupling these effects to understand their individual contributions to stress evolution remains a challenge. Determining whether this pressure recovery is predominantly due to swelling driven by unloading and subsequent resaturation, or to thermal expansion resulting from hydration heat or external thermal inputs, is essential for improving tunnel design, maintenance planning, and safety evaluations.

Field measurements revealed that the radial pressures acting on the tunnel support varied significantly (0.5–1.5 MPa) and continued to increase over the four-year monitoring period \citep{ZIEGLER-Arash2022}. These pressures, which are considerably higher than those recorded in the old Belchen Tunnel, align with the laboratory-measured swelling potential of OPA and suggest a strong temperature effect induced by construction and annual air temperature cycles. For instance, field measurements revealed that during the setting process of the gap grout and cast-in-place inner lining, temperature fluctuations resulted in pressure spikes. Additionally, a meaningful correlation was observed between the monitored radial total pressures and seasonal air temperature during tunnel operation.

As an intermediate material between the tunnel lining and host formation (OPA), a relatively soft bi-component grout was found to play a pivotal role in the system's thermal dynamics \citep{ziegler2017,ziegler2018}. Although grout can retain heat, the temperature within the tunnel exhibits considerable seasonal variation, typically being elevated in winter and reduced in summer. This seasonal fluctuation is strongly influenced by ventilation, which can introduce warm or cold air from the exterior into the tunnel cavity, resulting in oscillations in temperature.

The heat generated by tunnel traffic further complicates the thermal environment. These temperature variations can affect the structural integrity of the grout over time, particularly if it becomes excessively soft or loses stiffness owing to prolonged exposure to high temperatures. The low stiffness (compliance) of soft grout compared to stronger materials may lead to challenges in ensuring proper load transfer and stability, particularly in areas subjected to dynamic loads or temperature changes. 

The temperature evolution in the vicinity of the tunnel not only affects the mechanical behavior of the rock mass and leads to stress redistribution, but also amplifies the thermal expansion mismatch between the concrete lining and the surrounding rock, generating additional stresses \citep{Gens2009,Gens2010,Gens2013,manica2017,Toprak2017,tourchi_thermomechanical_2023}. Although this phenomenon has not been addressed in the literature, neglecting it during the design stage may threaten the structural integrity of the tunnel. Moreover, temperature variation in low-permeability rock (e.g., clay-rich shale) affects the pore-pressure field in the adjacent rock mass, thereby changing the effective stress and, correspondingly, the shear strength of the formation around the tunnel \citep{Monfared2011b,Mohajerani_2012,Monfared2014,thoeny2014}. This process can induce fracturing in the EDZ. Over time, these stresses can result in cracking, debonding, and deformation of both the tunnel lining and rock mass.

Conventional numerical modeling tools, although capable of simulating mechanical deformation and pore-pressure evolution, often neglect or oversimplify the thermal component. When thermal effects are included, they are typically imposed as boundary conditions without fully accounting for the coupling with deformation and fluid transport. In contrast, fully coupled THM models, such as those implemented in the \texttt{CODE\_BRIGHT} software \citep{olivella_numerical_1996}, offer a more rigorous approach for simulating the complex interactions that govern the behavior of tunnels in claystones. However, such advanced models require careful calibration, robust numerical stability, and validation against field-scale data. 

The material properties of the OPA, such as the thermal expansion coefficients, permeability anisotropy, and parameters for plastic flow, are highly variable and sensitive to spatial location and stress-path history \citep{Armand2004,bossart_characteristics_2011}. Moreover, the evolution of the EDZ, which governs many boundary and transition zones in the model, is inherently nonlinear and is poorly constrained. Thus, bridging the gap between theoretical modeling and practical application of STB remains a challenge.

The primary objective of this study is to develop a comprehensive THM model of the New Belchen Tunnel, building on our previous observational study \citep{ZIEGLER-Arash2022}, which presented and discussed field data, and to introduce a simplified thermo-mechanical (TM) model as a baseline for comparison. While earlier efforts primarily focused on in situ monitoring and initial observations, this study enhances the understanding of tunnel behavior through detailed numerical simulations, enabling a deeper exploration of coupled THM processes. 

This study simulated the immediate and long-term effects of temperature fluctuations, pore pressure evolution, and mechanical stress redistribution on the tunnel lining and adjacent rock mass behavior. The analysis quantifies the influence of temperature during the construction and operational phases on the structural integrity of the tunnel lining in the OPA shale. This approach provides new insights into improving tunnel design and maintenance strategies in similarly challenging geological environments. Furthermore, the insights developed are not limited to civil infrastructure; they are applicable to deep geological repositories (DGRs), where similar processes of heat generation, clay swelling, and long-term stress redistribution dominate \citep{gens2007,Wang2016,Bossart2017-2,Garitte2017b,Garitte2017a,Marschall2017,Muller2017,Birkholzer2018,Kim2024,Lanyon2024,Graupner2025,song2025longterm}. Integrating thermally coupled simulations into early stage design processes is indispensable for the robust prediction of system evolution over operational and post-closure lifetimes.

\section{Description of the new Belchen Tunnel}\label{sec_Description}

\begin{figure}[H]
    \centering
    \includegraphics[width=0.9\linewidth]{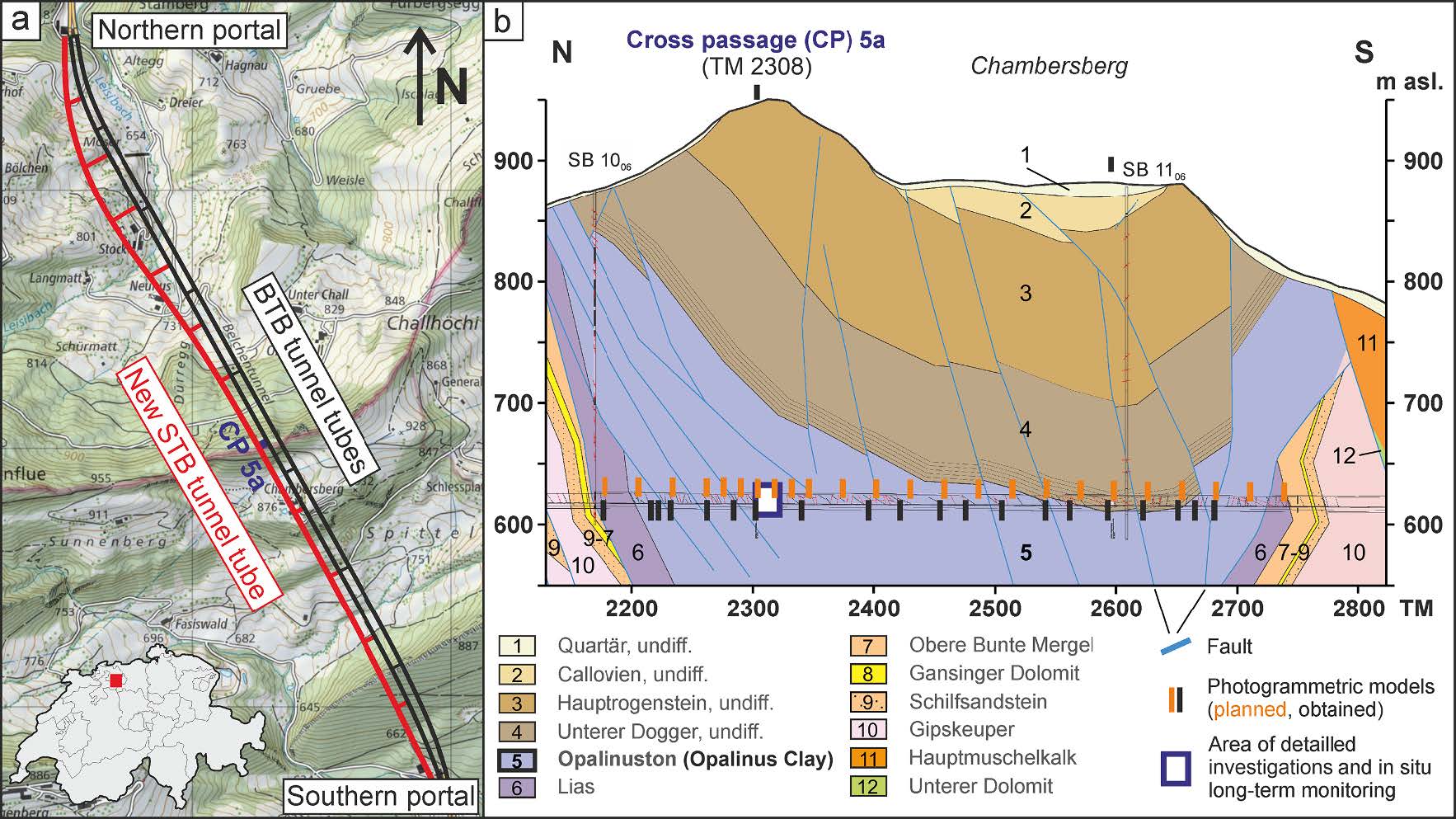}    
    \caption{(a) Map view of the Belchen tunnel trajectories and location within Switzerland. (b) Expected geological cross-section between TM 2130 and 2820 north of the New Belchen Tunnel (\citealt{ZIEGLER-Arash2022}).}
    \label{fig:Map}
\end{figure}

\begin{figure}[H]
    \centering
    \includegraphics[width=0.9\linewidth]{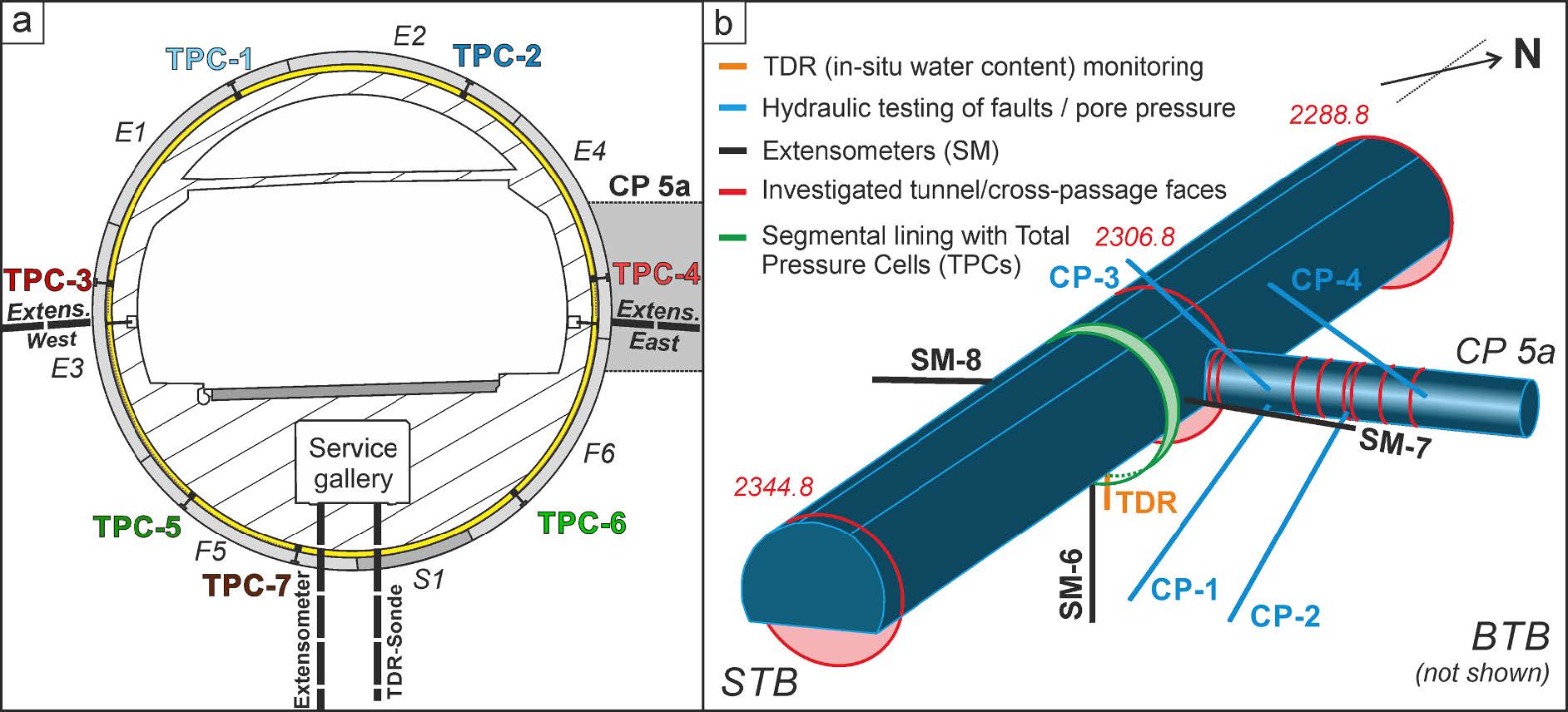}
    \caption{(a) Simplified 2D sketch of the monitoring section at TM 2317 with TPC-1–7, TDR borehole sonde in the tunnel invert, and three borehole extensometers. (b) 3D sketch of the monitoring section near cross-passage 5a (\citealt{ZIEGLER-Arash2022}).}
    \label{fig:2D_sketch}
\end{figure}

This section summarizes the geological setting, design features, and construction sequence of the New Belchen Tunnel, based primarily on the project documentation and monitoring study \citep{ziegler2017,ziegler2018,renz2019ensi,ZIEGLER-Arash2022,ziegler2023final}, unless otherwise noted. The New Belchen Tunnel (\textit{Sanierungstunnel Belchen}, STB) is an infrastructure project designed to address recurring maintenance issues associated with the aging Belchen Tunnel tubes originally excavated in the 1960s. Located in the Swiss Jura Mountains (\cref{fig:Map,fig:2D_sketch}), the existing tunnels suffered significant damage due to the swelling behavior of the surrounding anhydrite-rich marls (Gipskeuper) and OPA shale, which caused severe deformation of the tunnel linings. The new tunnel, constructed between 2016 and 2017, was designed to address these challenges by implementing modern tunneling methods and advanced support structures.

The STB is situated in a geologically complex area, traversing the folded and faulted formations of the Jura Mountain range, which includes Middle Triassic to Middle Jurassic sedimentary rocks. The geological conditions were well documented before construction owing to the tunnel's proximity to the existing Belchen tunnel tubes (located 40–116 m laterally) and additional exploratory drillings conducted in 2006 and 2009. The tunnel route passes through the Upper Triassic Gipskeuper and Aalenian to Toarcian OPA formations, both of which are known for their significant swelling potential.

The OPA shale, in particular, poses substantial challenges during tunnel construction. This formation, which is part of the Chambersburg syncline, is present in two distinct sections along the tunnel route. The swelling behavior of shale is exacerbated by its mineralogical composition, which includes a high proportion of sheet silicates, particularly illite/smectite interstratified clay minerals, which are known for their propensity to absorb water and expand. Geological complexity is further increased by the variable thickness of the OPA shale formation and the presence of steeply dipping faults and bedding planes, which cause frequent face instabilities during TBM excavation.

Excavation used a single-shield TBM approximately 75\,m long, with a 10\,m shield. This method limited disturbance to the surrounding rock compared with drill-and-blast excavation and thereby reduced the extent of the EDZ. The designed excavation diameter was 13.97\,m. 

The tunnel support system was carefully engineered to withstand high swelling pressures from surrounding rock and groundwater. The primary support consists of a 35 cm-thick outer lining made of precast concrete segments arranged in rings with seven segments per ring. These segments are complemented by a drainage membrane and a 65\,cm-thick inner cast-in-place concrete lining (see ~\cref{fig:2D_sketch}a). The construction of the inner lining was staged to manage the stresses imposed on the structure: the invert (the bottom portion of the tunnel) was concreted approximately 375 m behind the TBM cutting face, corresponding to approximately 30 days after excavation of that section. In contrast, the sidewalls and crown support were completed approximately 1.15 km behind the cutting face, corresponding to approximately 140 days after cutting the face.

A crucial aspect of tunnel design is managing the annular gap between the excavated rock and segmental lining. This gap was intentionally designed to be 22.5\,cm wide and was filled with a specially formulated two-component grout. The grout mixture, consisting of 90–95\% cement–bentonite water, was injected immediately after segment installation to provide structural support and flexibility. This grout has a uniaxial compressive strength ranging from 1.7 to 3.2 MPa \citep{antonioli2018}, and its mechanical properties were selected to accommodate the deformation behavior of the swelling clay shale. The gap grout reduced flow relative to an open annulus but remained much more permeable than the rock mass and worked together with the installed drainage measures. To reduce water-pressure build-up on the tunnel lining and hinder the migration of drained formation water along the tunnel into swellable formations, drainage slots and additional radial and longitudinal drainage membranes were installed at critical contacts between permeable and impermeable rock formations.

Extensive instrumentation was installed to monitor the tunnel's performance, particularly around tunnel meter (TM) 2317, where the overburden reached 325 m. This key monitoring section was equipped with radial total pressure cells (TPCs), borehole extensometers, and time-domain reflectometry (TDR) probes to measure mechanical pressure, radial deformation, and water content in the rock mass, respectively (see~\cref{fig:2D_sketch}a,b).

\section{Characteristics of OPA shale}\label{sec_Characteristics}

The OPA shale is a stiff, overconsolidated clay of the Lower Aalenian age (Middle Jurassic). It occurs in the Jura Mountains of northern Switzerland and the South German Jurassic period. The mineralogy consists mainly of sheet silicates (illite--smectite mixed layers, chlorites, and kaolinites), framework silicates (albite and K-feldspar), carbonates (calcite, dolomite, ankerite, and siderite), and quartz \citep{Bossart2002}. Three slightly different facies with distinct mineral proportions are commonly distinguished: a shaly facies in the lower part of the deposit, a \,\(\sim\)15\,m thick sandy--silty facies in the centre, and a sandy facies interstratified with the shaly facies in the upper part of the deposit. The clay mineral content ranges from 40\,

The clay was deposited under marine conditions; its pore water is highly mineralized (total dissolved salts up to 20 g/L) and contains a significant residual seawater component. The total OPA thickness was approximately 160\,m. Because OPA is the selected host rock for nuclear waste repositories in Switzerland, its behavior has been intensively studied through laboratory and in situ programs at the MT URL. In the STB monitoring section, the encountered facies were predominantly shaly \citep{ZIEGLER-Arash2022}.

A synthesis of the main physical and geotechnical parameters obtained from laboratory and in situ measurements on the OPA at the MT URL was reported by \citet{bossart_characteristics_2011}. Based on this information, the reference values and ranges used in our study are listed in \cref{tab:Reference_parameters}. Several parameters exhibit anisotropy with bedding; where only a single value is reported, it reflects limited data availability. From a soil-mechanics perspective, OPA is an indurated clay with relatively low porosity and low natural permeability; in petrophysical terms, however, its porosity is not unusually low for a claystone. Its rock strength is also low.

\begin{table}[H]
\centering
\caption{Reference parameters for OPA shale. The source codes in the final column are defined below the table.}
\label{tab:Reference_parameters}
\footnotesize
\renewcommand{\arraystretch}{1.2}
\begin{tabular}{@{}p{3.0cm}p{3.6cm}p{1.7cm}p{1.8cm}p{2.0cm}p{1.4cm}@{}}
\toprule
\shortstack[l]{\textbf{Property}\\\textbf{Category}} & \textbf{Parameter} & \textbf{Orientation\textsuperscript{a}} & \textbf{Reference Value} & \textbf{Range} & \textbf{Source} \\
\midrule

\multirow{3}{3.5cm}{\textbf{Mineralogy}} 
& Clay content (\%) & --- & 62 & 44--80 & A,B \\
& Carbonate content (\%) & --- & 14 & 5--22 & A,B \\
& Quartz content (\%) & --- & 18 & 10--20 & A,B \\
\midrule

\multirow{3}{3.5cm}{\textbf{Petrophysical Properties}} 
& Density, $\rho$ (\si{\gram\per\cubic\centi\meter}) & --- & 2.45 & $\pm$0.03 & A,B \\
& Water content, $w$ (\%) & --- & 6.1 & $\pm$1.9 & A,B \\
& Porosity (Water-loss), $\phi$ (\%) & --- & 16.2 & $\pm$2.2 & A,B \\
\midrule

\multirow{13}{3.5cm}{\textbf{Mechanical Properties}} 
& \multirow{2}{4cm}{Uniaxial compression strength, $R_{\mathrm{c}}$ (\si{\mega\pascal})} 
& Parallel & 10 & 4–17 & A,B \\
& & Perpendicular & 25.6 & 23.1–28.1 & A,B \\
\cmidrule{2-6}
& \multirow{2}{4cm}{Tensile strength, $R_{\mathrm{t}}$ (\si{\mega\pascal})} 
& Parallel & 2.0 & --- & A \\
& & Perpendicular & 1 & --- & A \\
\cmidrule{2-6}
& \multirow{2}{4cm}{Elastic modulus, $E$ (\si{\mega\pascal})} 
& Parallel & 7200 & 6300–8100 & A,B,D \\
& & Perpendicular & 2800 & 2100–3500 & A,B,D \\
\cmidrule{2-6}
& \multirow{2}{4cm}{Poisson's ratio, $\nu$} 
& Parallel & 0.24 & --- & A,B \\
& & Perpendicular & 0.33 & --- & A,B \\
\cmidrule{2-6}
& \multirow{2}{4cm}{Shear strength parameter, $c'$ (\si{\mega\pascal})} 
& Parallel & 5 & --- & A \\
& & Perpendicular & 2.2 & --- & A \\
\cmidrule{2-6}
& \multirow{2}{4cm}{Friction angle, $\varphi'$ (\si{\degree})} 
& Parallel & 25 & --- & A \\
& & Perpendicular & 16.5 & --- & A \\
\midrule

\multirow{4}{3.5cm}{\textbf{Thermal \& Thermo-mechanical Properties}} 
& \multirow{2}{4cm}{Thermal conductivity, $\lambda_{tc}$ (\si{\watt\per\meter\per\kelvin})} 
& Parallel & 2.1 & $\pm$10\% & B,C \\
& & Perpendicular & 1.2 & $\pm$10\% & B,C \\
\cmidrule{2-6}
& Heat capacity (dry, \SI{20}{\celsius}), $c_{\mathrm{s}}$ (\si{\joule\per\kilogram\per\kelvin}) & --- & 860 & $\pm$10\% & B,C \\
& Linear thermal expansion coefficient, $\alpha$ (\si{\per\kelvin}) & --- & $2.6 \times 10^{-5}$ & --- & C \\
\midrule

\multirow{2}{3.5cm}{\textbf{Hydraulic \& Hydro-mechanical Properties}} 
& Water permeability, $k$ (\si{\meter\per\second}) & --- & $10^{-13}$ & $10^{-12}$--$10^{-14}$ & A--C \\
& Biot's coefficient, $B$ & --- & 0.6 & 0.42--0.78 & C,D \\

\bottomrule
\end{tabular}
\begin{flushleft}
\footnotesize
\textsuperscript{a} Orientation relative to bedding plane.
\textbf{A}: \citet{bossart_characteristics_2011};
\textbf{B}: \citet{bossart2017};
\textbf{C}: \citet{wileveau2005thm};
\textbf{D}: \citet{wild_experimental_2018_part1}.
\end{flushleft}
\end{table}

At the STB location (\cref{fig:2D_sketch}), the overburden is approximately 325\,m, and the bedding dips are approximately $45^\circ$. No in-situ stress measurements were performed at the STB. We therefore used stress information from the MT URL at a comparable depth in the folded Jura, compiled by \citet{Martin2003} and updated in a more recent MT synthesis \citep{Jaeggi2018}, to define plausible starting values for the model. 

The major principal stress at the MT URL is subvertical and close to the overburden stress. Estimates of the intermediate and minimum principal stresses are based on, among other evidence, hydraulic-fracturing tests and borehole-breakout assessments. These regional values are not transferable 1:1 to the STB. Because the Belchen monitoring programme did not include in-situ stress measurements, the values in \cref{tab:insitu_stress} are modelling assumptions rather than site measurements or stress-calibration results.

\begin{table}[H]
    \centering
    \caption{Estimated in situ stress system (after \citet{Martin2003}).}
    \begin{tabular}{lllll}
        \hline
        \textbf{Principal stress} & \textbf{Orientation} & \textbf{Azimuth ($^\circ$)} & \textbf{Dip ($^\circ$)} & \textbf{Value (MPa)} \\
        \hline
        Major, $\sigma_1$        & Subvertical   & N210 & 70 & 6.0–7.0 \\
        \hline
        Intermediate, $\sigma_2$ & Subhorizontal & N320 & 10 & 4.0–5.0 \\
        \hline
        Minor, $\sigma_3$        & Subhorizontal & N50  & 20 & 2.0–3.0 \\
        \hline
    \end{tabular}
    \label{tab:insitu_stress}
\end{table}

\section{Theoretical formulation}\label{sec_thermoplasticity}
\subsection{THM formulation and balance equations}\label{sec_temperature_changes}

\noindent The theoretical THM formulation used for the simulation of the STB is a particular case of the general formulation presented in \citet{olivella_1994} for saturated and unsaturated media. For space reasons, only an outline is provided here. Two phases are considered: solid (s) and liquid (l), and two species are present: mineral (in the solid phase) and water (w). Solving a coupled THM problem requires the simultaneous solution of the following balance equations; hereafter, scalar, vector, and tensor quantities are denoted using regular, bold lowercase, and bold uppercase notation, respectively:

\noindent \textbf{Balance of solid:}
\begin{equation}
\frac{\partial}{\partial t}\!\left[\rho_s (1-\phi)\right] + \nabla \cdot \mathbf{j}_s = 0
\end{equation}

\noindent \textbf{Balance of water mass:}
\begin{equation}
\frac{\partial}{\partial t}\left(\rho_l \phi\right) + \nabla \cdot \mathbf{j}_l = f^w
\end{equation}

\noindent \textbf{Balance of internal energy:}
\begin{equation}
\frac{\partial}{\partial t}\!\left[E_s \rho_s (1-\phi) + E_l \rho_l S_l \phi\right]
+ \nabla \cdot \left(\mathbf{i}_c + \mathbf{j}_{Es} + \mathbf{j}_{El}\right) = f^Q
\end{equation}

\noindent \textbf{Equilibrium:}
\begin{equation}
\nabla \cdot \boldsymbol{\sigma} + \mathbf{b} = 0
\end{equation}

\noindent where $\phi$ is porosity, $\rho$ is density, $\mathbf{j}$ is the total mass flux, $\mathbf{u}$ is the solid displacement vector, $\boldsymbol{\sigma}$ is the stress tensor, $\mathbf{b}$ is the body force vector, $E$ is the specific internal energy, $\mathbf{i}_c$ is the conductive heat flux, and $\mathbf{j}_E$ is the energy flux due to mass motion.

Using the definition of the material derivative:
\begin{equation}
\frac{D_s (\cdot)}{Dt} 
= \frac{\partial (\cdot)}{\partial t} 
+ \frac{\mathrm{d}\mathbf{u}}{\mathrm{d}t} \cdot \nabla (\cdot)
\end{equation}

equation (1) becomes
\begin{equation}
\frac{D_s \phi}{Dt} 
= \frac{1}{\rho_s} \left[ (1 - \phi) \frac{D_s \rho_s}{Dt} \right] 
+ (1 - \phi) \nabla \cdot \frac{\mathrm{d}\mathbf{u}}{\mathrm{d}t}
\end{equation}

The solid mass balance can be eliminated by incorporating it into the water mass balance. 
Using the material derivative definition again, the following equation is obtained:
\begin{equation}
\phi \frac{D_s \rho_w}{Dt} 
+ \frac{\rho_w}{\rho_s} (1-\phi) \frac{D_s \rho_s}{Dt} 
+ \rho_w \nabla \cdot \frac{\mathrm{d}\mathbf{u}}{\mathrm{d}t} 
+ \nabla(\rho_w q_1 ) = 0 
\end{equation}

The first two derivatives of this expression can be further developed by
considering the dependence of the liquid and solid densities on temperature
and pore pressure, as follows:

\begin{equation}
\begin{cases}
    \rho_w = \rho_{w0} \exp\!\left[\beta_w (p_1 - p_{10}) 
    + b_w (T - T_{\text{ref}})\right] \\
    \rho_s = \rho_{s0} \exp\!\left[\beta_s (p_1 - p_{10}) 
    + 3 b_s (T - T_{\text{ref}})\right]
\end{cases}
\end{equation}

\noindent where $\beta_w$ and $\beta_s$ are the water and solid compressibilities, respectively, and  $b_w$ and $b_s$ are the volumetric and linear thermal expansion coefficients for water and the solid grain, respectively. Expanding the first two derivatives of Equation (8) results in

\begin{equation}
\left[ \phi b_w + (1 - \phi) 3b_s \right] \frac{D_s T}{Dt}
+ \phi \beta_w \frac{D_s p_w}{Dt}
+ (1 - \phi) \beta_s \frac{D_s p_s}{Dt}
+ \nabla \cdot \frac{d\mathbf{u}}{dt}
+ \frac{\nabla \cdot (\rho_w \mathbf{q}_1)}{\rho_w}
= 0
\end{equation}

Equation (9) contains the THM couplings that explain the variation in pore pressure when a temperature change is applied to the claystone. The first term expresses the differential thermal expansion of the solid and liquid phases, respectively. The second and third terms represent the volume changes in the water and solid phases associated with the pore pressure change. The fourth term indicates the volume change of the material skeleton (including contributions from stress, pore pressure, and temperature), and the fifth term is the volume change associated with the flow of water in or out of the element considered. The pore pressure generated is the result of the interplay of all these terms in each particular case.

The formulation was completed using various constitutive laws that describe the phenomena under consideration. These have been presented and discussed elsewhere (e.g., \citealp{gens2017}). The main ones correspond to the flow of heat by conduction, advective flow of water, and mechanical constitutive law for mechanical behavior.

Heat conduction is governed by Fourier's law:

\begin{equation}
\mathbf{i}_c = -\lambda_{\text{tc}} \nabla T
\end{equation}

\noindent where $\lambda_{\text{tc}}$ is the coefficient of thermal conductivity. Water flow is controlled by Darcy's law:

\begin{equation}
\mathbf{q}_1
= -\frac{\mathbf{k}}{\mu_l} \left( \nabla p_1 - \rho_w \mathbf{g} \right)
\end{equation}

\noindent
where $\mathbf{k}$ is the intrinsic permeability tensor (with principal
components $k_1$ and $k_2$; see \cref{tab:model_parameters}), and $\mu_l$ is the dynamic 
viscosity of water.


\subsection{Thermomechanical model for argillaceous hard soils–weak rocks\label{sec_Thermomechanical_model}}

This section describes a constitutive model aimed at reproducing the main features of the THM behavior of OPA shale. In particular, it incorporates the non-isothermal behavior of strength and stiffness, nonlinear isotropic hardening before peak strength, post-peak softening, a non-associated flow rule, time-dependent deformation, and permeability dependence on accumulated plastic deformation. The constitutive model employed for the simulations was developed and implemented in \texttt{CODE\_BRIGHT} by \citet{manica2017} and \citet{tourchi_thermomechanical_2023}.

At the continuum scale, the formulation follows a phenomenological strategy. Each major response observed in laboratory tests is represented by a dedicated constitutive component rather than derived from an explicit mineral-scale or crack-scale model. Cross-anisotropic thermoelasticity describes reversible deformation; a non-isothermal Mohr--Coulomb surface with nonlinear hardening, softening, and non-associated flow describes instantaneous irreversible deformation; an overstress law describes deviatoric creep; and accumulated plastic deformation controls the intrinsic permeability. These components share the same stress and state variables and are integrated with the coupled balance equations. Their parameters are calibrated first against triaxial and creep tests and are then carried into the tunnel analysis. This approach retains the main observed couplings with a tractable parameter set, but it does not resolve discrete cracking, mineral-scale swelling, or interface damage \citep{manica2017,tourchi_thermomechanical_2023}.

The numerical integration procedure is summarized here because the model combines temperature-dependent elasticity, plastic hardening and softening, and viscoplastic flow within each coupled time increment. In \texttt{CODE\_BRIGHT} terminology, SPA denotes the stress point algorithm used to integrate these constitutive equations at each Gauss point \citep{vaunat2000spa}. For a global increment, the procedure first evaluates an elastic trial stress. A trial state inside the yield surface is accepted as elastic; otherwise, an implicit return-mapping step solves for the updated stress, plastic multiplier, and hardening variable while enforcing the flow rule and consistency condition. Local Newton iterations, with subincrementation where required, provide the updated stress and tangent for the global equilibrium iteration. SPA is therefore the local integration procedure, not an additional constitutive law.

The model is implemented in terms of effective stresses, adopting the following generalized expression that accounts for the effects of potential desaturation:
\begin{equation}
\boldsymbol{\sigma}' = \boldsymbol{\sigma} + S_e\, s\, B\, \mathbf{I}
\end{equation}
\noindent
where \(\boldsymbol{\sigma}'\) is the effective stress tensor, \(S_e\) is the effective degree of saturation, \(s\) is the suction, \(B\) is Biot's coefficient, and \(\mathbf{I}\) is the identity tensor.


\subsubsection{Thermoelastic components}

The relationship between the elastic modulus (\(E\)) and temperature is characterized by the following logarithmic function:
\begin{equation}
E(T) = E(T_0) \left[ 1 - \gamma_E \ln\!\left( \frac{T}{T_0} \right) \right]
\end{equation}
\noindent
where \(\gamma_E\) controls the rate of reduction in the Young's modulus with temperature.

Assuming that the coefficient of thermal expansion is independent of the stresses, the hypoelastic strain increment is defined as the sum of the thermal and mechanical components:
\begin{equation}
d\boldsymbol{\varepsilon}^{e} = d\boldsymbol{\varepsilon}^{(e,\sigma)} + \frac{1}{3}\, d\varepsilon_v^{(e,T)}\, \mathbf{I}
\end{equation}

\noindent
where $d\boldsymbol{\varepsilon}^{(e,\sigma)}$ is the increment of elastic strain caused by changes in the effective stress, and $d\varepsilon_v^{(e,T)}$ is the elastic volumetric strain increment caused by changes in temperature $T$. The latter can be defined as:

\begin{equation}
d\varepsilon_v^{(e,T)} = 3 \alpha\, dT
\end{equation}

\noindent
where $\alpha$ is the linear thermal expansion coefficient of the medium, which depends on the material mineralogy, temperature, and pressure changes, although it can be assumed to be constant for practical purposes.


\subsubsection{Thermoplastic components}

\noindent The total plastic strain increment in the thermoplastic constitutive model is formulated as the sum of instantaneous plastic deformations and time-dependent viscoplastic deformations, capturing both immediate and progressive responses of the OPA shale under thermal and mechanical loading 

\begin{equation}
\mathrm{d} \boldsymbol{\varepsilon} = \mathrm{d} \boldsymbol{\varepsilon}^{\mathrm{ep}} + \dot{\boldsymbol{\varepsilon}}^{\mathrm{vp}} \, \mathrm{d}t
\end{equation}

\noindent
where $d\boldsymbol{\varepsilon}$ is the total strain increment, consisting of the elastic-plastic strain increment $d\boldsymbol{\varepsilon}^{\mathrm{ep}}$ and the viscoplastic strain increment $\dot{\boldsymbol{\varepsilon}}^{\mathrm{vp}}\, dt$. 

The viscoplastic strain rate tensor is computed as follows:
\begin{equation}
\dot{\boldsymbol{\varepsilon}}^{\mathrm{vp}} = \frac{3}{2} \frac{\dot{\varepsilon}_{\mathrm{eq}}^{\mathrm{vp}}}{q} \mathbf{s}
\label{eq:creep_tensor}
\end{equation}

\noindent
where $q$ is the deviatoric stress invariant defined as

\begin{equation}
q = \left( \frac{3}{2} \mathbf{s} : \mathbf{s} \right)^{1/2}
\label{eq:deviatoric_q}
\end{equation}

The scalar viscoplastic strain rate is expressed as
\begin{equation}
\dot{\varepsilon}_{\mathrm{eq}}^{\mathrm{vp}}
= \gamma_{\text{vp}} \left\langle q - \sigma_s \right\rangle^n
  \left( 1 - \varepsilon_{\mathrm{eq}}^{\mathrm{vp}} \right)^m
\label{eq:creep_scalar}
\end{equation}

\noindent
where $\gamma_{\text{vp}}$ is the viscosity parameter, $\sigma_s$ is the viscoplastic threshold stress, 
$\langle \cdot \rangle$ denotes the Macaulay brackets, $n$ and $m$ are material parameters, 
and $\varepsilon_{\mathrm{eq}}^{\mathrm{vp}}$ is the equivalent viscoplastic strain, defined as
\begin{equation}
\varepsilon_{\mathrm{eq}}^{\mathrm{vp}}(t) = 
\int_0^{t} \left( \frac{2}{3}\,
\dot{\boldsymbol{\varepsilon}}^{\mathrm{vp}} : 
\dot{\boldsymbol{\varepsilon}}^{\mathrm{vp}} \right)^{1/2} \,\mathrm{d}t 
\label{eq:creep_accum}
\end{equation}

At each time step, the deviatoric invariant $q$ is evaluated from the current effective stress. The Macaulay brackets in \cref{eq:creep_scalar} suppress creep when $q\leq\sigma_s$. Above this threshold, the rate increases with the overstress $q-\sigma_s$ according to exponent $n$, while the factor involving the accumulated viscoplastic strain and exponent $m$ progressively reduces the transient rate. The factor $3/2$ in \cref{eq:creep_tensor} makes the tensorial rate consistent with the equivalent-strain definition in \cref{eq:creep_accum}. The flow is coaxial with the deviatoric stress and has zero trace; the law therefore produces shear creep but no time-dependent volumetric strain. The parameters $\gamma_{\mathrm{vp}}$, $\sigma_s$, $n$, and $m$ are held constant in the tunnel simulations. Temperature affects creep only indirectly through the temperature-dependent stiffness, strength, and evolving stress state; the creep-rate parameters have no explicit temperature, saturation, or suction dependence \citep{manica2017}.


\subsubsection{Yield surface}
 
Under low deviatoric stresses, the response is linearly elastic and is characterized by a transversely isotropic form of Hooke's law. For a higher deviatoric stress, plastic deformations develop upon reaching the yield surface, which is characterized by the generalized non-isothermal Mohr-Coulomb criterion as follows:

\begin{equation}
f^{T} = \left( \cos \theta + \frac{1}{\sqrt{3}} \sin \theta \sin \varphi_{\text{mob}}^{T} \right) J 
- \sin \varphi_{\text{mob}}^{T} \left( c_{\text{peak}}^{T_0} \cot \varphi_{\text{mob}}^{T} + p \right)
\end{equation}

\noindent where $\varphi_{\text{mob}}$ is the mobilized friction angle, $c_{\text{mob}}$ is the mobilized cohesion, and the remaining variables are stress invariants with their usual definitions. The friction angle varies in a piecewise manner as follows:

\begin{equation}
\varphi_{\text{mob}}^{T} =
\begin{cases}
\varphi_{\text{ini}} + \dfrac{\epsilon_{eq}^p}
{a_{\text{hard}} - \dfrac{\epsilon_{eq}^p \left( a_{\text{hard}} - \xi_1 \right)}{\xi_1}} 
& \text{if } \epsilon_{eq}^p \leq \xi_1 \\[10pt]

\varphi_{\text{peak}} 
& \text{if } \xi_1 < \epsilon_{eq}^p \leq \xi_2 \\[10pt]

\varphi_{\text{peak}} - 
\dfrac{ \epsilon_{eq}^p - \xi_2 }
{ a_{\text{soft}} - 
\dfrac{ a_{\text{soft}} - (\xi_3 - \xi_2) }
     { (\varphi_{\text{peak}} - \varphi_{\text{res}}) }
\dfrac{ \epsilon_{eq}^p - \xi_2 }{ \xi_3 - \xi_2 } }
& \text{if } \xi_2 < \epsilon_{eq}^p \leq \xi_3 \\[10pt]

\varphi_{\text{res}} 
& \text{if } \epsilon_{eq}^p > \xi_3
\end{cases}
\end{equation}

\noindent where \( \varphi_{\text{ini}} \) is the initial friction angle,
\( \varphi_{\text{peak}} \) is the peak friction angle,
\( \varphi_{\text{res}} \) is the residual friction angle,
\( \varepsilon_{eq}^p \) is the state variable controlling hardening/softening,
\( \xi_1 \) is the value of \( \varepsilon_{eq}^p \) at peak strength,
\( \xi_2 \) is the value of \( \varepsilon_{eq}^p \) at which softening begins,
\( \xi_3 \) is the value of \( \varepsilon_{eq}^p \) at which the residual strength is reached,
\( a_{\text{hard}} \) is a parameter controlling the curvature of the function in the hardening branch,
and \( a_{\text{soft}} \) is a parameter controlling the curvature of the function in the softening branch.
The equivalent plastic strain is defined as
\begin{equation}
\varepsilon_{eq}^p =
\left( \frac{2}{3}\,
\boldsymbol{\varepsilon}^p : \boldsymbol{\varepsilon}^p \right)^{1/2}.
\end{equation}
where \(\boldsymbol{\varepsilon}^p\) is the plastic strain tensor. The comma in the original version of this equation was a typographical error and has been removed; the revised equation uses the standard colon for tensor double contraction.
The strength dependence on temperature is incorporated through the parameters
\( \varphi \) and \( c \). In the case of the friction angle, the following equations
were adopted to define \( \varphi_{\text{ini}} \), \( \varphi_{\text{peak}} \),
\( \varphi_{\text{res}} \) as a function of temperature:

\begin{equation}
\varphi_{\text{ini}}(T) = \varphi_{\text{ini}}^{(T_0)} 
\left[ 1 - \mu_\varphi \ln\left( \frac{T}{T_0} \right) \right]
\end{equation}

\begin{equation}
\varphi_{\text{peak}}(T) = \varphi_{\text{peak}}^{(T_0)} 
\left[ 1 - \mu_\varphi \ln\left( \frac{T}{T_0} \right) \right]
\end{equation}

\begin{equation}
\varphi_{\text{res}}(T) = \varphi_{\text{res}}^{(T_0)} 
\left[ 1 - \mu_\varphi \ln\left( \frac{T}{T_0} \right) \right]
\end{equation}

\noindent where \( \varphi_{\text{ini}}^{(T_0)} \), \( \varphi_{\text{peak}}^{(T_0)} \), and \( \varphi_{\text{res}}^{(T_0)} \) are the initial, peak, and residual friction angles, respectively, at the reference temperature \( T_0 \), and \( \mu_\varphi \) is a parameter controlling the rate of change of the friction angle with temperature. As in the case of mechanical loading, it is assumed that the Mohr-Coulomb envelope rotates around a fixed point; therefore, we can define the temperature-dependent mobilized cohesion as

\begin{equation}
c_{\text{mob}}(T) = c_{\text{peak}}^{(T_0)} 
\cot\left( \varphi_{\text{peak}}^{(T_0)} \right) 
\tan\left( \varphi_{\text{mob}}^{(T)} \right)
\end{equation}

\noindent
where $c_{\text{peak}}^{(T_0)}$ is the peak cohesion at reference temperature $T_0$.

\begin{figure}[H]
    \centering
    \includegraphics[width=0.85\linewidth]{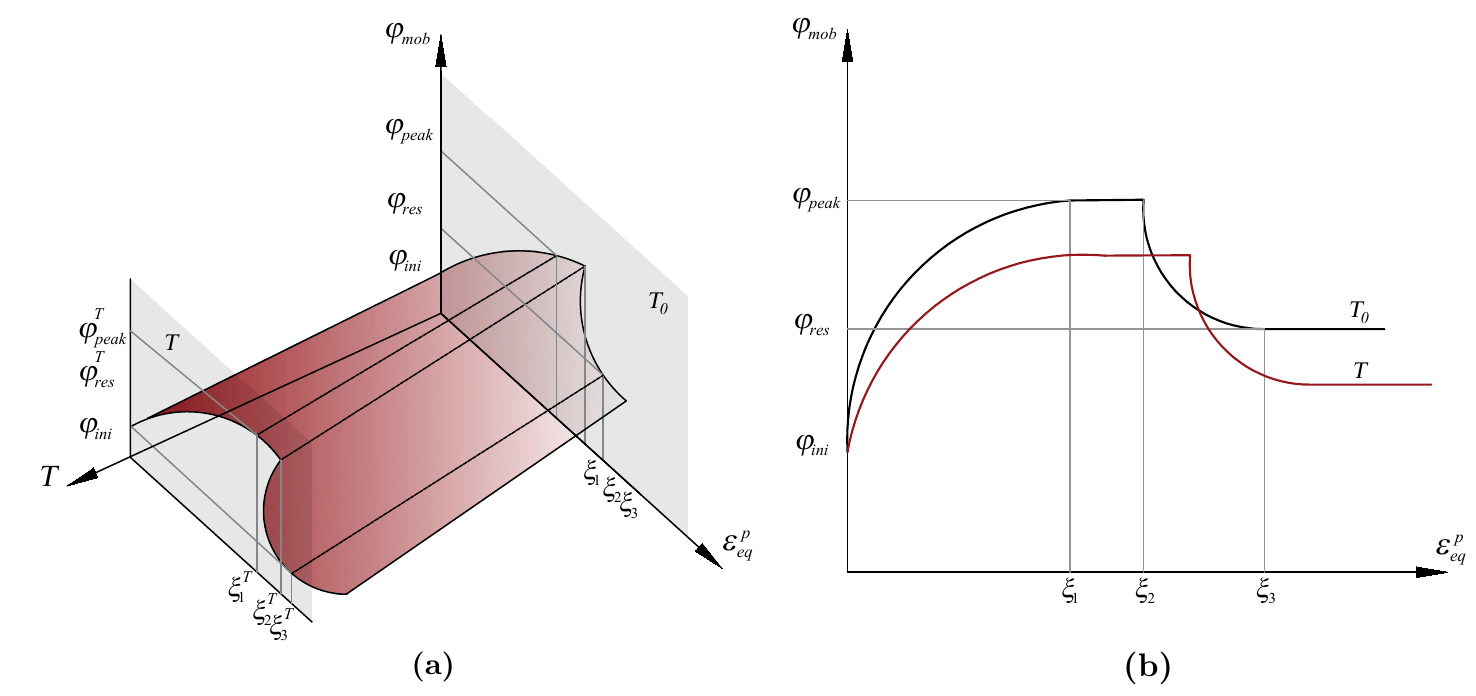}
    \caption{(a) Evolution of the yield envelope; (b) Mobilized friction angle as a function of temperature and \( \epsilon_{eq}^p \).}
    \label{fig:hardening_softening}
\end{figure}

\Cref{fig:hardening_softening} presents the mobilized friction angle 
$\varphi_{\text{mob}}$ as a function of temperature $T$ and equivalent plastic
strain $\varepsilon_{eq}^p$. This 3D representation shows how the friction
angles $\varphi_{\text{ini}}$, $\varphi_{\text{peak}}$, and $\varphi_{\text{res}}$
govern this evolution. At the initial temperature ($T_0$) and zero plastic
strain, the mobilized friction angle is $\varphi_{\text{mob}} = \varphi_{\text{ini}}$. 
As temperature and plastic strain increase, $\varphi_{\text{mob}}$ rises towards
$\varphi_{\text{peak}}$ and, after peak, decreases and gradually approaches the
residual friction angle $\varphi_{\text{res}}$.

\subsubsection{Flow rule }

\noindent A non-associated flow rule is adopted. Rather than defining a separate
plastic potential, we derive the flow rule directly from the yield function
\(f(p,J_2,\theta)\):
\begin{equation}
\frac{\partial g}{\partial \boldsymbol{\sigma}}
= \;\omega\,\frac{\partial f}{\partial p}\,\frac{\partial p}{\partial \boldsymbol{\sigma}}
\;+\;     \frac{\partial f}{\partial J_2}\,\frac{\partial J_2}{\partial \boldsymbol{\sigma}}
\;+\;     \frac{\partial f}{\partial \theta}\,\frac{\partial \theta}{\partial \boldsymbol{\sigma}} .
\label{eq:flowrule}
\end{equation}

\noindent
Here \(g\) is the plastic potential, and \(0\le\omega\le1\) controls only the volumetric
component of plastic flow (the term with \(p\)): \(\omega=1\) recovers the associated
rule; \(\omega=0\) gives isochoric plastic flow.

The increment of the plastic strain $\mathrm{d}\boldsymbol{\varepsilon}^p$ is defined as
\begin{equation}
\mathrm{d}\boldsymbol{\varepsilon}^p = \mathrm{d}\Lambda \, \frac{\partial g}{\partial \boldsymbol{\sigma}}
\end{equation}

\noindent
where $\mathrm{d}\Lambda$ is the plastic multiplier increment, defined as

\begin{equation}
\mathrm{d}\Lambda = \frac{1}{H} \left(
    \frac{\partial f}{\partial \boldsymbol{\sigma}} \, d\boldsymbol{\sigma}
    + \frac{\partial f}{\partial \varphi_{\text{mob}}}
      \frac{\partial \varphi_{\text{mob}}}{\partial T} \, \mathrm{d}T
\right)
\end{equation}

The hardening modulus $H$ is expressed as
\begin{equation}
H = -\frac{\partial f}{\partial \varphi_{\text{mob}}}
    \frac{\partial \varphi_{\text{mob}}}{\partial \varepsilon_{eq}^p}
    \frac{\partial \varepsilon_{eq}^p}{\partial \boldsymbol{\varepsilon}^p}
    \cdot \frac{\partial g}{\partial \boldsymbol{\sigma}}
\end{equation}

Therefore, plastic deformation is also affected by the thermal variation of the strength parameters. The model also accounts for cross-anisotropy (or transverse isotropy) through a non-uniform scaling of the stress tensor (\citealp{manica2016}):

\begin{equation}
\boldsymbol{\sigma}^{\text{ani}} =
\begin{bmatrix}
    \dfrac{\sigma_{11}^r}{c_N} & c_S \sigma_{12}^r & \sigma_{13}^r \\
    c_S \sigma_{12}^r & c_N \sigma_{22}^r & c_S \sigma_{23}^r \\
    \sigma_{13}^r & c_S \sigma_{23}^r & \dfrac{\sigma_{33}^r}{c_N}
\end{bmatrix}
\label{eq:anisotropic_stress}
\end{equation}

\noindent
where $c_N$ and $c_S$ are the normal and shear scaling factors, respectively, used to represent the anisotropy of the material, and $\sigma_{ij}^r$ are the components of the reference stress tensor $\boldsymbol{\sigma}^r$.


\subsubsection{Hydro-mechanical components}

\noindent
The equivalent degree of saturation is given by the following form of the expression proposed by \citet{vangenuchten1980}:

\begin{equation}
S_e = \frac{S_1 - S_{r1}}{S_{ls} - S_{r1}} = 
\left[ 1 + \left( \frac{p_g - p_1}{P} \right)^{\frac{1}{1 - m_w}} \right]^{-n_w}
\end{equation}

\noindent
where $S_1$ is the degree of saturation, $S_{r1}$ is the residual degree of saturation, and $S_{ls}$ is the degree of saturation under fully saturated conditions (typically $S_{ls} = 1$). 
$p_g$ and $p_1$ are the gas and liquid pressures, respectively; $m_w$ and $n_w$ are van Genuchten shape parameters controlling the curvature of the water retention function; and $P$ is the air-entry pressure.

The model assumes that the intrinsic permeability evolves with plastic deformations to capture the increase in permeability due to damage in the OPA shale. The plastic multiplier $\Lambda$ characterizes the accumulated plastic deformations, and the intrinsic permeability tensor $\mathbf{k}$ is defined as

\begin{equation}
\mathbf{k} = \begin{cases} \mathbf{k}_0 \exp\left[ \eta(\Lambda - \Lambda_{\text{thr}}) \right] & \text{if } \Lambda > \Lambda_{\text{thr}} \\ \mathbf{k}_0 & \text{if } \Lambda \leq \Lambda_{\text{thr}} \end{cases}
\end{equation}

\noindent where $\mathbf{k}_0$ is the intrinsic permeability tensor of the intact rock, $\eta$ is a scalar parameter controlling the rate of permeability increase, and $\Lambda_{\text{thr}}$ is the threshold plastic multiplier, beyond which permeability increases.


\section{Numerical simulation }
\subsection{Simulation of the mechanical behavior of OPA shale}

The numerical simulations presented in this study leverage extensive experience in THM modeling of argillaceous formations, particularly in the context of underground excavations and nuclear waste storage applications \citep{tourchi2019coupled,tourchi2020thm,tourchi_thermomechanical_2023}. The modeling approach builds on previous work on full-scale tunnel simulations in Callovo-Oxfordian claystone, which shares characteristics with the Opalinus Clay encountered at the Belchen tunnel site.

The constitutive model parameters were obtained by calibration against temperature-controlled triaxial compression tests and long-term creep tests on OPA shale \citep{Zhang2007}. Cylindrical specimens were isotropically consolidated to a confining stress of $\sigma_3=3\,\mathrm{MPa}$ and then heated at $2\text{–}5\,^\circ\mathrm{C}\!/\mathrm{h}$ to target temperatures between $20^\circ\mathrm{C}$ and $115^\circ\mathrm{C}$, during which axial and radial thermal strains were recorded. After thermal equilibrium, axial loading was applied at a constant strain rate of $1\times10^{-7}\,\mathrm{s}^{-1}$ to failure under undrained conditions on specimens cored with bedding inclinations of $30^\circ\text{–}40^\circ$. The volumetric strains observed during the undrained tests result primarily from thermal expansion of the solid skeleton and pore fluid, as well as the compressibility of the partially saturated pore system, rather than from drainage-induced volume change.

\begin{figure}[H]
    \centering
    \includegraphics[scale=1]{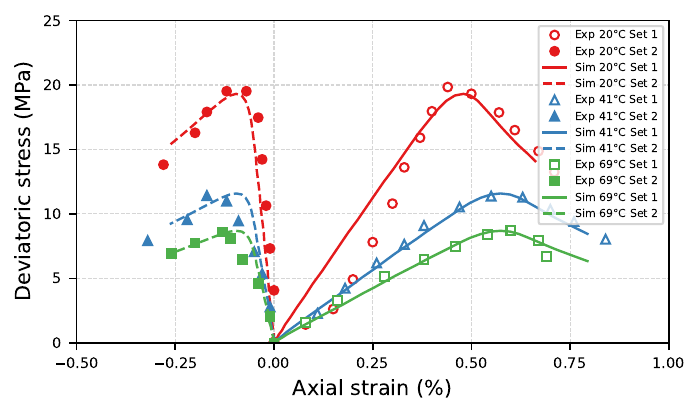}
    \caption{Stress--strain curves in triaxial tests on OPA shale. \textit{Experimental observations} (Exp; \citealp{Zhang2007}) and \textit{model simulations} (Sim). Set~1 and Set~2 identify the paired experimental and simulated series available at each temperature.}
    \label{fig:stress_strain_curves}
\end{figure}

\begin{figure}[H]
    \centering
    \includegraphics[scale=1]{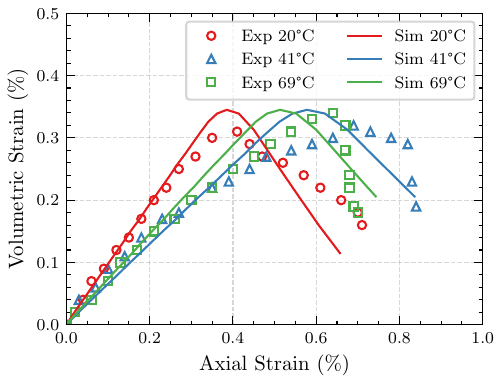}
    \caption{Volume change in triaxial tests on OPA shale. \textit{Experimental observations} (Exp; \citealp{Zhang2007}) and \textit{model simulations} (Sim).}
    \label{fig:volumetric_strains}
\end{figure}

These laboratory procedures were replicated numerically as full finite-element boundary-value simulations in \texttt{CODE\_BRIGHT}, using quadrilateral meshes with single-point Gaussian integration and adaptive time-stepping. Identical thermal, hydraulic, and mechanical boundary conditions were applied, including undrained conditions during the shearing phase. As shown in ~\cref{fig:stress_strain_curves}, the simulated and experimental deviatoric stress–strain curves are compared, and ~\cref{fig:volumetric_strains} shows that the model reproduces the laboratory data with deviations within 5\% for all loading paths.

\begin{figure}[H]
    \centering
    \includegraphics[scale=0.9]{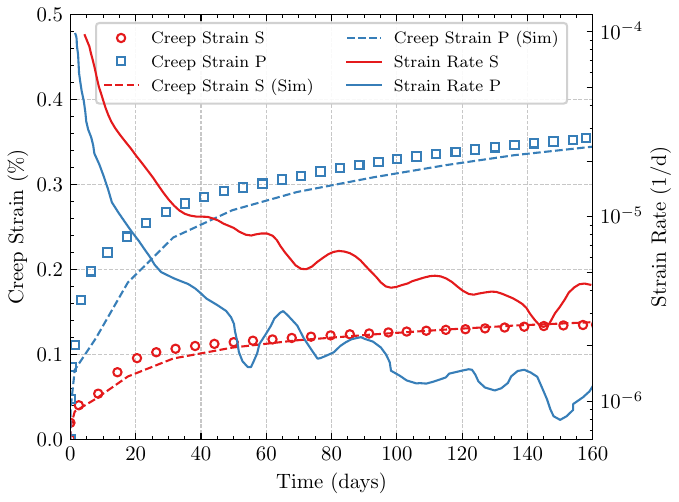}
    \caption{Creep tests on OPA shale. \textit{Experimental observations} (Exp; \citealp{naumann_experimental_2007}) and \textit{model simulations} (Sim).}
    \label{fig:Creep_tests}
\end{figure}

To investigate the time-dependent behavior of the OPA shale, creep tests were conducted under triaxial loading conditions (e.g., \citet{naumann_experimental_2007}; see also \citet{GOTZEN2025}). In these tests, the axial displacements were halted once the target deviatoric stress was reached, and the stress state was held constant for a specified duration. The experiments were performed at a constant confining pressure of 12 MPa, with triaxial loads increasing in stages. Each stage lasted up to 200 days, and the differential stress was increased in 2 MPa increments, starting from 13 MPa. The goal was to compare the creep behaviors for different bedding orientations relative to the direction of the principal stress. Samples loaded parallel to the bedding (P-samples) showed markedly different transient creep strains from those loaded normal to the bedding (S-samples). \Cref{fig:Creep_tests} shows the time-dependent deformations. The parameters governing the time-dependent response are listed in \cref{tab:model_parameters}. Recent fully saturated and consolidated tests reported secondary creep rates several times faster than those in partially saturated OPA, indicating that the creep potential may have been underestimated in earlier studies \citep{GOTZEN2025}.

The creep anisotropy is captured through the model's transversely isotropic elastic stiffness and the non-uniform stress scaling in \cref{eq:anisotropic_stress}. The scaling acts before the stress enters the yield function and viscoplastic flow rule, allowing the scalar creep law to produce an orientation-dependent response.

\begin{table}[H]
  \centering
  \caption{Constitutive model parameters for triaxial, creep, and THM simulations.}
  \label{tab:model_parameters}
  \begin{tabular}{llll}
    \toprule
    \multicolumn{4}{c}{\textbf{Instantaneous (elastic and strength)}} \\
    \midrule
    Parameter                            & Symbol                    & Value                 & Unit      \\
    \midrule
    Elastic modulus (parallel)           & $E_\parallel$             & 7200                  & MPa       \\
    Elastic modulus (perpendicular)      & $E_\perp$                 & 2800                  & MPa       \\
    Poisson’s ratio                      & $\nu_\parallel,\nu_\perp$ & 0.2                   & --        \\
    Peak friction angle                  & $\varphi_{\mathrm{peak}}$ & 22                    & $^\circ$  \\
    Initial friction angle               & $\varphi_{\mathrm{ini}}$  & 9.35                  & $^\circ$  \\
    Residual friction angle              & $\varphi_{\mathrm{res}}$  & 14.74                 & $^\circ$  \\
    Peak cohesion                        & $c_{\mathrm{peak}}$       & 3.5                   & MPa       \\
    Hardening exponent                   & $a_{\mathrm{hard}}$       & 0.001                 & --        \\
    Softening exponent                   & $a_{\mathrm{soft}}$       & 0.05                  & --        \\
    Plastic flow weight                  & $\omega$                  & 0.6                   & --        \\
    \midrule
    \multicolumn{4}{c}{\textbf{Time-dependent (viscoplastic)}} \\
    \midrule
    Threshold stress for creep onset     & $\sigma_s$                & 4.79                  & MPa       \\
    Viscoplastic rate coefficient        & $\gamma_{\mathrm{vp}}$    & $9.69 \times 10^{-6}$ & MPa$^{-n}$ day$^{-1}$\\
    Hardening exponent                   & $n$                       & 4.98                  & --        \\
    Softening exponent                   & $m$                       & 20                    & --        \\
    \midrule
\multicolumn{4}{c}{\textbf{Hydraulic (van Genuchten) and damage}} \\
\midrule
Intrinsic permeability (horizontal) & $k_1$                     & $1\times10^{-20}$     & m$^2$     \\
Intrinsic permeability (vertical)   & $k_2$                     & $2\times10^{-20}$     & m$^2$     \\
Retention curve parameter            & $A$                       & 1                     & --        \\
van Genuchten shape parameter        & $m_w$                     & 3                     & --        \\
van Genuchten shape parameter        & $n_w$                     & 0.33                  & --        \\
Air-entry pressure                   & $P$                       & 14.3                  & MPa       \\
Damage threshold (plastic multiplier)& $\Lambda_{\mathrm{thr}}$    & $1\times10^{-5}$      & --        \\
Damage amplification factor          & $\eta$                    & 7.5                   & --        \\

    \midrule
    \multicolumn{4}{c}{\textbf{Thermal (temperature dependence)}} \\
    \midrule
    Friction–temperature coupling        & $\mu_{\varphi}$           & 0.17                  & --        \\
    Elastic-modulus--temperature coupling & $\gamma_E$                & 0.34                  & --        \\
    \bottomrule
  \end{tabular}
\end{table}


\subsection{Aspects of analysis and material characteristics}

\begin{figure}[H]
    \centering
    \includegraphics[width=1\linewidth]{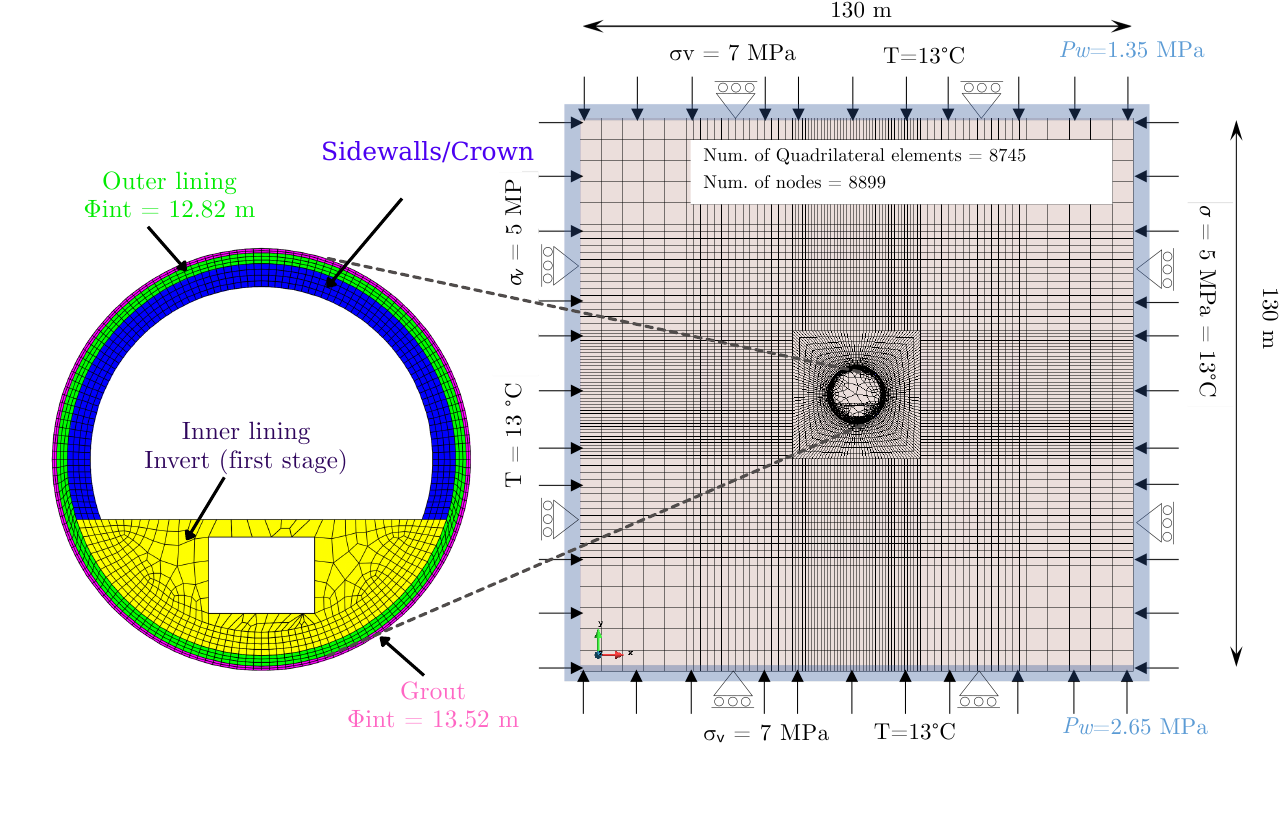}
    \caption{Geometry, meshing, and initial and boundary conditions.}
    \label{fig:model_mesh}
\end{figure}

To interpret the observed THM behavior of the STB, numerical simulations were
performed using the finite element code \texttt{CODE--BRIGHT}
\citep{olivella_1994,olivella_numerical_1996}. The model domain and finite
element mesh are depicted in \cref{fig:model_mesh}, where the initial stress
state can be noted. The initial vertical (maximum) and horizontal principal stresses
have been assumed to be \(7 \, \text{MPa}\) and \(5 \, \text{MPa}\) at the
tunnel level, respectively (see \cref{tab:insitu_stress}). Because no in situ
stress tests were carried out at the New Belchen Tunnel (STB), the far-field
stresses used in the simulations were established by combining an overburden
calculation with regional priors from the Mont Terri Underground Rock
Laboratory at a comparable burial depth in the folded Jura
\citep{Martin2003}. 

The vertical stress was estimated as
\(\sigma_v = \gamma_{\text{avg}} H\), with tunnel-axis depth
\(H = 325\,\mathrm{m}\) and a depth-averaged unit weight
\(\gamma_{\text{avg}} \approx 21.5\,\mathrm{kN\,m^{-3}}\) for the
limestone–marl–OPA succession, giving a total vertical stress
\(\sigma_v \approx 7\,\mathrm{MPa}\). The horizontal stress followed the
assumed at-rest earth-pressure ratio in total stresses,
\(K_{0,\text{tot}} \approx 0.7\), yielding
\(\sigma_h = K_{0,\text{tot}} \sigma_v \approx 5\,\mathrm{MPa}\); therefore,
the major principal stress was subvertical and the intermediate subhorizontal,
consistent with the regional stress field. The adopted unit weight represents the
logged sedimentary succession, while \(K_{0,\text{tot}}\) is a regional modelling
assumption. Neither value was measured at the monitoring section or adjusted during
calibration. In the THM formulation, the
effective stresses are then obtained from these total stresses via
\(\boldsymbol{\sigma}' = \boldsymbol{\sigma} - B\,p\,\mathbf{I}\), where \(p\)
is the pore pressure. Thus, \(K_0\) is interpreted here in terms of total
stresses, while all constitutive calculations are performed in effective
stresses.

A hydrostatic pore-pressure field was used for initialization and maintained at the external boundary. It was anchored at \(p_{\text{top}}=1.35\,\mathrm{MPa}\) and increased linearly with depth to \(p_{\text{bottom}}=2.65\,\mathrm{MPa}\) over the \(130\,\mathrm{m}\) model height. No background pore-pressure measurements were available at the Belchen monitoring section, so this profile is an idealized initial and far-field condition rather than a site-derived pressure distribution. The baseline stresses and hydraulic heads were retained in all analyses; the calibration concerned the material and thermal parameters rather than the far-field state.

The dimensions of the model were \(130 \, \text{m} \times 130 \, \text{m}\). The distance from the tunnel wall to the boundary was \(58 \, \text{m}\). The model also incorporates the geometry of the liner and the gap grout between the liner and OPA shale (\cref{fig:model_mesh}). The mesh contained \(8745\) quadrilateral elements and \(8899\) nodes, and was refined near the gallery and tunnel to effectively deal with the higher temperature and pore pressure gradients in this zone. An initial constant temperature of \(13^\circ \text{C}\) was assumed throughout the geometry. 

Mechanically, the OPA, gap grout, segmental lining, and inner lining were represented by conforming continuum domains with shared interface nodes. Displacement continuity was therefore enforced at the lining--grout and grout--rock boundaries, corresponding to a fully bonded idealization. The model contains no contact, frictional sliding, cohesive debonding, or gap-opening law. Hydraulically, the lining--grout assembly was set to impermeable and the rock-facing tunnel boundary was set to no flow. This deliberate sectional idealization was adopted because the gap-grout permeability was unavailable and the waterproofing membrane redirects much of the actual drainage longitudinally. Such out-of-plane drainage cannot be represented in a 2D plane-strain section. Allowing radial flow through the lining and grout while permitting dissipation only at the distant external boundary would therefore misrepresent the field drainage path. Drainage membranes, pipes, and drainage towards cross-passages were not represented explicitly. The computed near-field pressures must consequently be read as the response under this idealized no-flow condition, rather than as a direct prediction of field pore pressure. The monitoring data show strongly nonuniform load transfer and are consistent with local gap-grout cracking, but they do not provide direct evidence of interface slip or loss of contact \citep{ZIEGLER-Arash2022}. Interface damage should therefore be viewed as a possible field mechanism outside the present model, not as an outcome calculated here.

The measured temperature variations at discrete sensor locations along the tunnel lining (e.g., TPC-1 to TPC-7) were used to define a spatially continuous thermal boundary condition using bilinear interpolation. Rather than assigning a uniform temperature history across the entire boundary, the measured data were grouped according to their position (crown, sidewalls, invert), and separate time-dependent thermal loads were constructed for each region of the tunnel. This allowed the simulation to capture both the temporal evolution and spatial heterogeneity of thermal inputs owing to localized processes such as concrete hydration (in the invert and sidewalls), ventilation-driven air temperature fluctuations (in the crown), and operational thermal disturbances. The resulting boundary conditions ensured a realistic approximation of the distributed thermal environment experienced by the tunnel lining, thereby improving the model’s ability to reproduce the coupled thermomechanical response observed in the field. The interpolation also smoothens the discontinuities between the sensor points, mitigating unrealistic thermal gradients and ensuring numerical stability.

In the thermal--mechanical model, the invert concrete was idealized as a uniform cast reaching the springline (mid-height) and extending continuously along the monitored ring. This simplified construction geometry does not resolve local variations in the as-built pour elevation. Moreover, the potential thermal interaction with the existing Belchen Tunnel (BTB), located east of the STB at a distance of approximately two tunnel diameters, was not explicitly represented. Given the separation and intervening rock mass, conductive diffusion over the time scales analyzed (order of one year) is expected to be limited (diffusion length $L\!\approx\!\sqrt{\alpha t}\!\sim\!5$\,m for $\alpha\!\sim\!7\times10^{-7}$\,m$^2$/s), and any BTB-related perturbation was small compared with the local curing and seasonal variations at the STB wall.

To account for the anisotropic thermal behavior of the OPA shale, direction-dependent thermal properties were implemented into the model. Specifically, the thermal conductivity was defined as $\lambda_{\parallel} = 2.1 \, \mathrm{W/(m \cdot K)}$ parallel to the bedding and $\lambda_{\perp} = 1.2 \, \mathrm{W/(m \cdot K)}$ perpendicular to the bedding, based on the ranges reported for Opalinus Clay in \cref{tab:Reference_parameters}. These values were used to construct a thermal conductivity tensor aligned with the bedding orientations. Anisotropic thermal conductivity was incorporated via a coordinate transformation consistent with the dip angle of the bedding planes ($45^\circ$). Similarly, the heat capacity was assigned with respect to the orientation, although its directional variation had only a minor influence on the results. This approach captured the directional differences in the computed thermal gradients shown below.

 The various stages of tunnel excavation were mimicked by the different phases of the numerical analysis, as shown in  \cref{tab:construction_sequence}.

\begin{table}[H]
\centering
\caption{Construction sequence of the New Belchen Tunnel at the monitoring section}
\label{tab:construction_sequence}
\begin{tabular}{llll}
\toprule
\textbf{Phase} & \textbf{Start Date} & \textbf{End Date} & \textbf{Duration} \\
\midrule
TBM excavation \& precast lining\textsuperscript{a}            & 09 Feb 2016   & 21 Jun 2017      & 16 months \\
Annular-gap grouting (behind TBM shield)\textsuperscript{a}    & Feb 2016      & Jun 2017          & 16 months \\
Invert \& service-tunnel segment concreting\textsuperscript{b} & 01 Nov 2016   & $\approx$ 22 Nov 2016 & $\sim$3 weeks \\
Cross-passage (CP~5a) excavation\textsuperscript{b}            & mid-Jan 2017  & mid-Jan 2017      & $\sim$1 week \\
Sidewalls \& crown concreting\textsuperscript{b}               & 08 Apr 2017   & 08 Apr 2017       & 1 day \\
\bottomrule
\end{tabular}

\vspace{2pt}
\begin{minipage}{0.97\linewidth}\footnotesize
\textbf{Notes:} \textsuperscript{a} Project-wide activity window (kept as in project documentation). 
\textsuperscript{b} Monitoring-section events verified in \citet{ZIEGLER-Arash2022}.
\end{minipage}
\end{table}

The STB was excavated from south to north, with rock mass structure orientations indicating a near east--west strike \citep{ZIEGLER-Arash2022}. The tunnel axis is approximately perpendicular to the bedding strike and intersects southward-dipping layers. Consequently, the tunnel advances through an anisotropic and heterogeneous rock mass, including bedding, joints, and tectonic faults, which significantly influences stress redistribution and deformation during and after excavation. The Opalinus Clay exhibits transversely isotropic behavior, with high contrasts in stiffness along and across bedding planes. Regarding elastic stiffness anisotropy, a ratio between the horizontal and vertical Young’s moduli of \(2.57\) was employed, consistent with the values reported by \citet{bossart_characteristics_2011}. For the present engineering idealization, the directional Young's modulus was estimated with an inverse quadratic, or $n=2$ Hankinson-type, interpolation between the bedding-normal and bedding-parallel end members \citep{bodig_jayne_1982,ting1996anisotropic}:

\begin{equation}
\frac{1}{E(\theta)} = \frac{\cos^2\theta}{E_{\perp}} + \frac{\sin^2\theta}{E_{\parallel}}
\label{eq:young}
\end{equation}

\noindent where $\theta$ is the angle between the axis of interest and the bedding normal. This relation is a two-end-member engineering interpolation, not a complete transversely isotropic compliance formulation; the latter would require the full set of anisotropic elastic constants \citep{ting1996anisotropic}. Assuming a bedding dip of $45^\circ$ towards the south and a tunnel advancing from south to north, it was used to calculate the longitudinal, vertical, and horizontal transverse moduli reported below.

Rock strength anisotropy was represented by two direction-dependent multipliers that act on the bedding-normal and bedding-parallel strength end members. We adopted \(c_{N}=1.33\) (normal to bedding) and \(c_{S}=1.0\) (parallel to bedding), which raise the across-bedding shear resistance relative to the along-bedding value, thereby favouring the characteristic EDZ seen when drifts are oriented approximately perpendicular to bedding strike. 

In particular, this choice reproduces (i) lateral, sidewall-focused extensional damage with shorter crown/invert zones and (ii) a horizontally elongated EDZ footprint (aspect \(\sim 2\!-\!3{:}1\)), both documented for Opalinus Clay in field mappings and numerical studies at Mont Terri \citep{Bossart2002,Martin2002,nussbaum2011analysis,Lisjak2015}. The multipliers act on the strength end members; orientation mixing gives \(c_{\mathrm{eff}}(\theta)\) and \(\varphi_{\mathrm{eff}}(\theta)\), while the stress-tensor scaling in \cref{eq:anisotropic_stress} accounts for the directional stress response.

The variation in the shear strength parameters with direction was approximated using a trigonometric interpolation approach. For a given angle $\theta$ measured from the bedding normal, the effective cohesion and friction angle were calculated using the following expressions:
\begin{equation}
    c_{\text{eff}}(\theta) = c_{\perp} \cos^2\theta + c_{\parallel} \sin^2\theta
    \label{eq:cohesion_direction}
\end{equation}

\begin{equation}
    \varphi_{\text{eff}}(\theta) = \varphi_{\perp} \cos^2\theta + \varphi_{\parallel} \sin^2\theta
    \label{eq:friction_direction}
\end{equation}

\noindent where $c_{\perp}$ and $\varphi_{\perp}$ represent the cohesion and friction angle perpendicular to bedding, and $c_{\parallel}$ and $\phi_{\parallel}$ are the corresponding values parallel to bedding. $\theta$ is the angle between the axis of interest and bedding normal. We adopted a representative bedding dip of $45^\circ$ to the south, consistent with structural measurements reported for the monitoring area (mean dip $54^\circ \pm 9^\circ$ toward $193^\circ$ at the STB face and $58^\circ \pm 20^\circ$ toward $168^\circ$ from borehole logs at CP~5a). Given this local variability due to folding and minor faulting, $45^\circ$ lies within the central range and provides a stable reference for directional projections. A sensitivity check varying the dip by $\pm 10^\circ$ changes $E(\theta)$ by less than $\sim 3\%$ and the interpolated $(c,\varphi)$ by less than $\sim 0.5^\circ$, without altering the predicted EDZ lobe orientation.

The interpolations in \cref{eq:young,eq:cohesion_direction,eq:friction_direction} assign stiffness and strength to the tunnel's longitudinal, vertical, and transverse axes from their angle to the bedding normal. In parallel, \cref{eq:anisotropic_stress} scales the stress components in the bedding-normal and bedding-parallel material axes before evaluating yielding and viscoplastic flow. The first step defines directional material parameters; the second modifies the stress measure used by the constitutive law. Together they provide a tractable 2D representation of bedding anisotropy, but they do not replace a full three-dimensional transversely isotropic formulation. Table~\ref{tab:model_parameters} reports the scalar parameters used for the laboratory calibration, whereas \cref{tab:Directional} reports the effective directional values assigned in the tunnel model after projection and anisotropic scaling. The two tables therefore describe different stages of parameter assignment rather than duplicate parameter sets.

\begin{table}[H]
\centering
\caption{Directional mechanical parameters of OPA shale along the tunnel axes used in the simulations.}
\label{tab:Directional}
\renewcommand{\arraystretch}{1.1}
\begin{tabular}{lccc}
\hline
\textbf{Tunnel axis} & \textbf{Young's modulus (MPa)} & \textbf{Cohesion (MPa)} & \textbf{Friction angle} \\
\hline
Longitudinal (axis)  & 4032 & 4.25 & $20.8^\circ$ \\
Vertical             & 4032 & 4.25 & $20.8^\circ$ \\
Horizontal (transverse) & 7200 & 4.90 & $24.9^\circ$ \\
\hline
\end{tabular}
\end{table}

The constitutive parameters adopted in the tunnel analyses are those calibrated in
\cref{tab:model_parameters}. Directional stiffness and strength for the
transversely isotropic OPA shale were prescribed according to
\cref{tab:Directional}, where the longitudinal, vertical, and horizontal
(transverse) axes are aligned with the tunnel axis, the gravity direction, and
in the in-plane transverse direction, respectively. Intrinsic permeability of the
intact rock, $k_0$, together with the parameter $\eta$ controlling its
nonlinear evolution with plastic straining, are also taken from
\cref{tab:model_parameters}. The material formulation includes time-dependent
viscoplastic deformation (creep) as described in Section~4.2.2, with parameters
obtained from triaxial creep calibration (\cref{tab:model_parameters}). This
mechanism governs long-term deformation under quasi-constant stress and
contributes to delayed stress redistribution around the excavation.

\begin{table}[H]
\centering
\caption{Thermo-mechanical properties for the segmental lining and the annular gap-grout (verified from \citealp{ZIEGLER-Arash2022}).}
\label{tab:corrected_lining_grout_properties}
\renewcommand{\arraystretch}{1.2}
\setlength{\tabcolsep}{2.8pt}
\begin{tabular}{p{4.6cm}p{5.0cm}p{4.0cm}p{2.6cm}}
\toprule
\textbf{Element} & \textbf{Parameter} & \textbf{Symbol (unit)} & \textbf{Value} \\
\midrule
\multirow{5}{=}{\textbf{Segmental lining} (concrete)} 
  & Young's modulus & $E$ (GPa) & 30 \\
  & Poisson's ratio & $\nu$ (--) & 0.20 \\
  & Coefficient of thermal expansion & $\alpha$ ($10^{-6}$ K$^{-1}$) & 33 \\
  & Thermal conductivity & $\lambda_{tc}$ (W\,m$^{-1}$\,K$^{-1}$) & 1.6 \\
  & Specific heat capacity & $c$ (J\,kg$^{-1}$\,K$^{-1}$) & 900 \\
\midrule
\multirow{7}{=}{\textbf{Two-component gap-grout} (cement--bentonite)} 
  & Young's modulus & $E$ (GPa) & 1.0–2.0 \\
  & Young's modulus (tangent)\textsuperscript{\ddag} & $E_{50}$ (GPa) & 0.6–1.7 \\
  & Poisson's ratio & $\nu$ (--) & 0.32 \\
  & Coefficient of thermal expansion & $\alpha$ ($10^{-6}$ K$^{-1}$) & 30 \\
  & Thermal conductivity & $\lambda_{tc}$ (W\,m$^{-1}$\,K$^{-1}$) & 0.6 \\
  & Specific heat capacity & $c$ (J\,kg$^{-1}$\,K$^{-1}$) & 910 \\
  & Uniaxial compressive strength & $f_c$ (MPa) & 1.7–3.2 (avg.\ $2.3 \pm 0.6$) \\
\bottomrule
\end{tabular}
\end{table}


\subsection{Results of the analysis}\label{sec:results}

\noindent
In this section, the numerical results are compared with the field observations. Because radial pressures are not direct outputs of \texttt{CODE--BRIGHT}, they were post-processed by projecting the Cauchy stress tensor $\boldsymbol{\sigma}$ from the closest Gauss point onto the local outward normal of the tunnel boundary. The radial stress at a sensor location was computed as
$\sigma_r=\mathbf{n}^{\mathsf T}\boldsymbol{\sigma}\,\mathbf{n}$,
where $\mathbf{n}$ is the unit normal at the position of each total pressure cell (TPC-1 to TPC-7). \Cref{fig:2D_sketch} indicates the TPC locations and the temperature instrumentation at the monitored section. Seven TPCs (TPC-1 to TPC-7) were mounted in recesses on the {outer} face of specially designed steel-reinforced lining segments and distributed around the full circumference. Time-varying {measured} temperatures from thermistors co-located with each TPC were imposed on the inner concrete surface as Dirichlet boundary conditions for the heat equation; the temperature field within the rock and grout then evolved by conduction and THM coupling. Details of the sensor layout and logging are provided by \citet{ZIEGLER-Arash2022}.

\begin{figure}[H]
    \centering
    \includegraphics[scale=1]{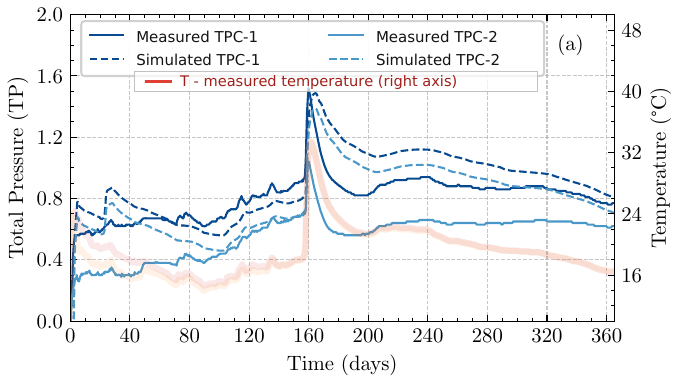}
    \includegraphics[scale=1]{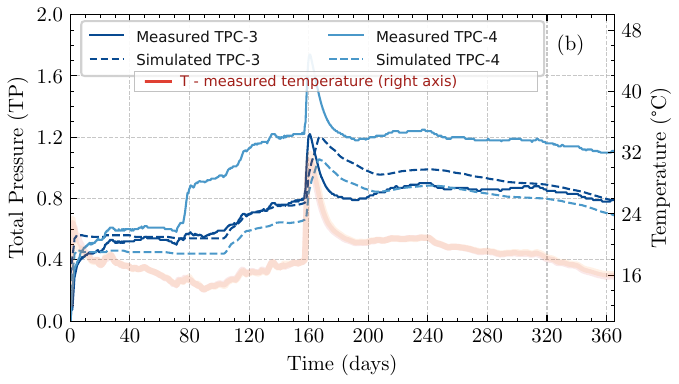}
    \includegraphics[scale=1]{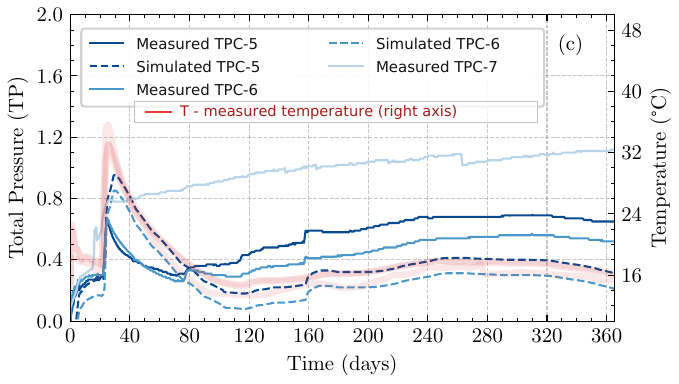}
     \caption{Evolution of total radial pressure (left axis; blue) and temperature, $T$ (right axis; red), at seven total pressure cells between 1~Nov~2016 and 1~Nov~2017. {(a)} Crown: TPC-1 and TPC-2. {(b)} Sidewalls: TPC-3 and TPC-4. {(c)} Invert: TPC-5, TPC-6, and TPC-7. Pressure observations are plotted with solid lines and model results with dashed lines. The red curves show measured $T$ at the corresponding TPC locations.}
    \label{fig:Total_Pressures}
\end{figure}

\paragraph{Crown (TPC-1, TPC-2).}
At the crown (\cref{fig:Total_Pressures}a), early installation and curing led to initial temperature and pressure increases; when the air and lining temperature was $\sim\!20^\circ$C above ambient, TPC-1 reached $\sim\!1.5$\,MPa and TPC-2 $\sim\!0.9$\,MPa. During operation, a clear seasonal co-variation of temperature and pressure persists, indicating that thermally induced expansion of the lining--rock system is the primary driver of the cyclic component.

\paragraph{Sidewalls (TPC-3, TPC-4).}
For the sidewalls (\cref{fig:Total_Pressures}b), rapid temperature increases are accompanied by abrupt pressure increases. At TPC-4, a pronounced peak occurred near day 75, which was temporally aligned with the excavation of cross-passage 5a. A second thermal peak ($\approx\!35^\circ$C) near day 150 is mirrored by a pressure peak at TPC-3 of approximately $ 1.2$ MPa, emphasizing the role of curing and staged segment placement.

\paragraph{Invert (TPC-5, TPC-6, TPC-7).}
The sensors in the invert showed a strong thermo-mechanical coupling (\cref{fig:Total_Pressures}c). Early curing-related spikes (up to $\sim 35\,^{\circ}\mathrm{C}$) coincide with sharp pressure increases. Subsequently, both temperature and pressure generally decrease as thermal conditions return towards ambient. The model underestimates the longer-term pressures at TPC-5 and TPC-6, while TPC-7 exhibits a sustained rise consistent with local swelling or load redistribution in the invert.

\medskip
\noindent
The coupled temperature–pressure analysis revealed that construction-induced thermal processes significantly impact the total pressure within the lining. The numerical model captures the main trends but locally underestimates the peaks linked to site-specific activities (e.g., cross-passage excavation) that are not represented in the 2D section. The close correspondence between the measured temperatures and observed pressures supports the need to include coupled thermo-mechanical effects during construction-phase analysis.

The integrated temperature--pressure analysis confirmed that construction-induced thermal processes substantially governed the evolution of the total pressure within the lining. For the thermal boundary condition, we prescribed the {measured} inner-lining temperature histories at the TPC locations as time-dependent Dirichlet conditions on the corresponding crown, sidewall, and invert arcs. Between the sensors, circumferential temperatures were obtained by angular interpolation, while axial uniformity was assumed in the 2D section; the far-field model boundaries were held at the initial rock temperature. This implementation ties the thermal input directly to the in-situ measurements at TM~2317 (see \citealt{ZIEGLER-Arash2022} for the sensor layout and calibration).

In the first weeks to months after installation and curing, the simulations reproduced the observed co-variations of temperature and total pressure at the crown (TPC-1, TPC-2) and sidewalls (TPC-3, TPC-4), including the magnitude and timing of the early peaks linked to hydration heat and staged segment placement. As construction progressed, systematic deviations emerged in specific sensors. At the sidewall, the rise near day 75 at TPC-4 coincided with excavation of cross-passage 5a, a local three-dimensional disturbance that was not represented in the present 2D model. The second thermal and pressure peak around day 150 at TPC-3 is reproduced reasonably well. At the invert, the model underestimates the long-term pressures at TPC-5 and TPC-6 and captures the initial thermoelastic response at TPC-7 but not its subsequent monotonic increase. These differences are consistent with longer-term mechanisms such as grout-property evolution and clay-shale swelling that amplify pressure even as temperatures decline (cf. \cref{fig:Total_Pressures}). 


\subsubsection{Temperature distributions (directional comparison)}

\begin{figure}[H]
  \centering
  \includegraphics[scale=0.95]{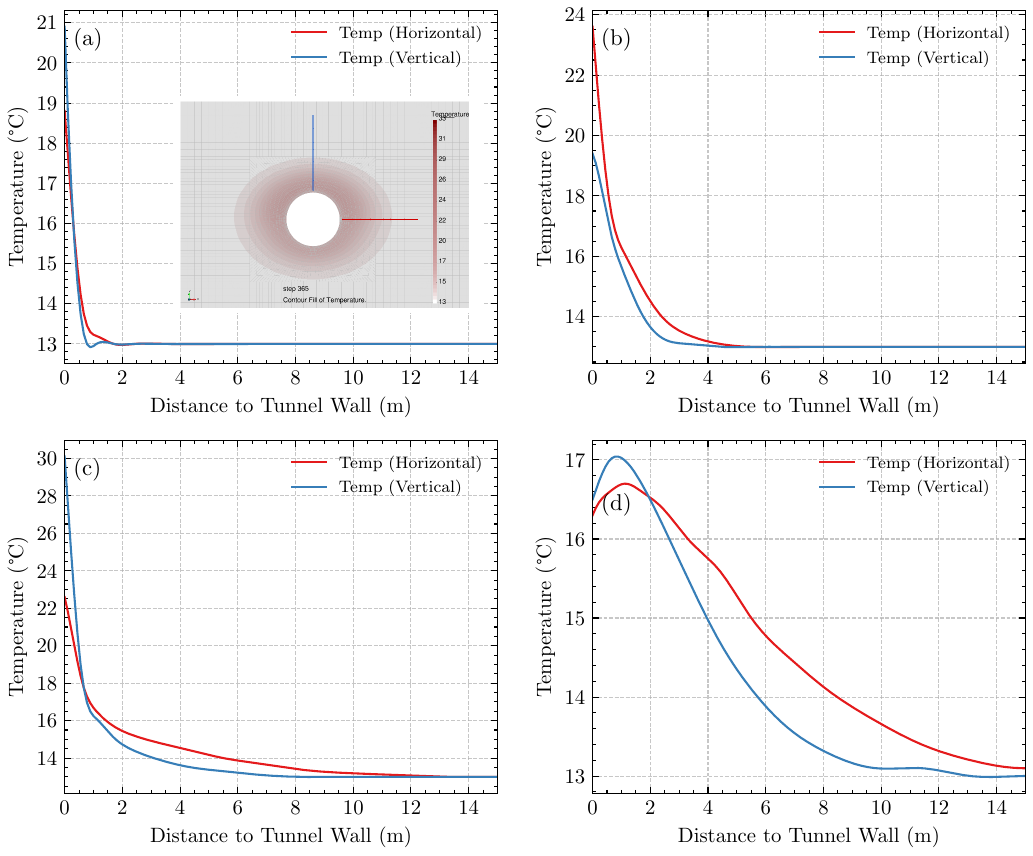}
  \caption{Computed temperature profiles along two orthogonal paths emanating from the wall: Direction~H (2D “horizontal”, red) and Direction~V (2D “vertical/oblique”, blue) at \text{(a)} $t_a=7$\,days, \text{(b)} $t_b=30$\,days, \text{(c)} $t_c=160$\,days, and \text{(d)} $t_d=365$\,days after lining installation. The grey inset in panel~(a) is a day-365 contour used only to show the two sampling directions. In all cases the temperature decreases with distance; the slower decay along Direction~H reflects the larger thermal conductivity in that direction ($\lambda_H>\lambda_V$).}
  \label{fig:temp_distr}
\end{figure}

\noindent
\Cref{fig:temp_distr}(a–d) compares computed temperature profiles along two orthogonal paths emanating from the wall: a lateral path (Direction~H; the 2D ``horizontal'' ray) and a vertical/oblique path (Direction~V; the 2D ``vertical/oblique'' ray). Profiles are shown at representative times $t_a=7$~days, $t_b=30$~days, $t_c=160$~days, and $t_d=365$~days after lining installation. In all cases, the temperature decreased with distance from the wall; however, the decay along Direction~H was systematically {slower} than that along Direction~V. This behavior is consistent with the thermal anisotropy adopted for the clay shale, with conductivity larger in Direction~H than in Direction~V, that is, $\lambda_H>\lambda_V$ (cf. \cref{tab:Reference_parameters}: $\lambda_H\!\approx\!2.1$ and $\lambda_V\!\approx\!1.2$~W\,m$^{-1}$\,K$^{-1}$). The higher-conductivity direction spreads heat more efficiently, producing a broader thermal plume and a shallower near-wall gradient; hence, the apparently slower decay along Direction~H.

\begin{figure}[H]
  \centering
  \begin{tabular}{cc}
    \includegraphics[width=0.45\linewidth,trim=0 58 0 0,clip]{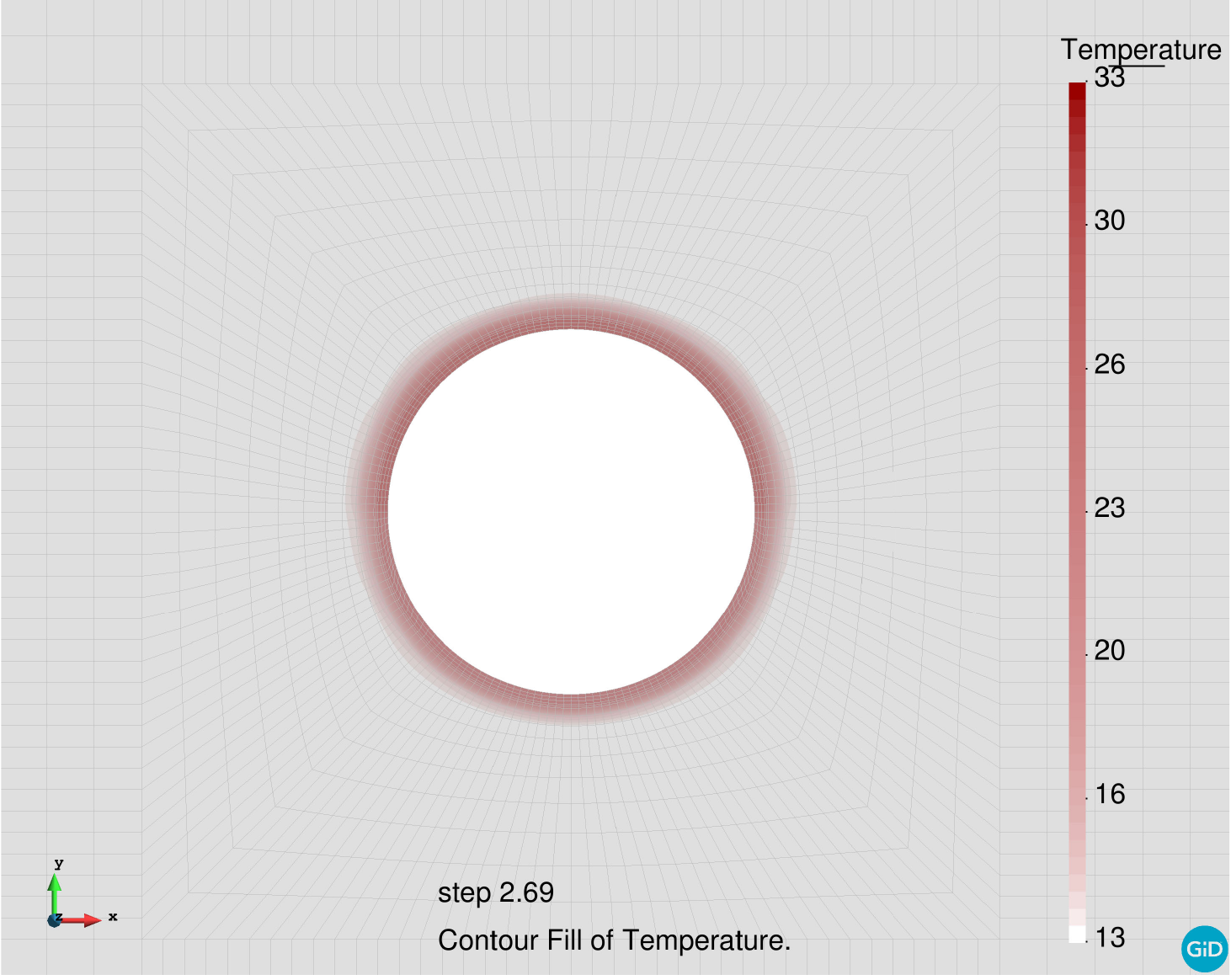} & \includegraphics[width=0.45\linewidth,trim=0 58 0 0,clip]{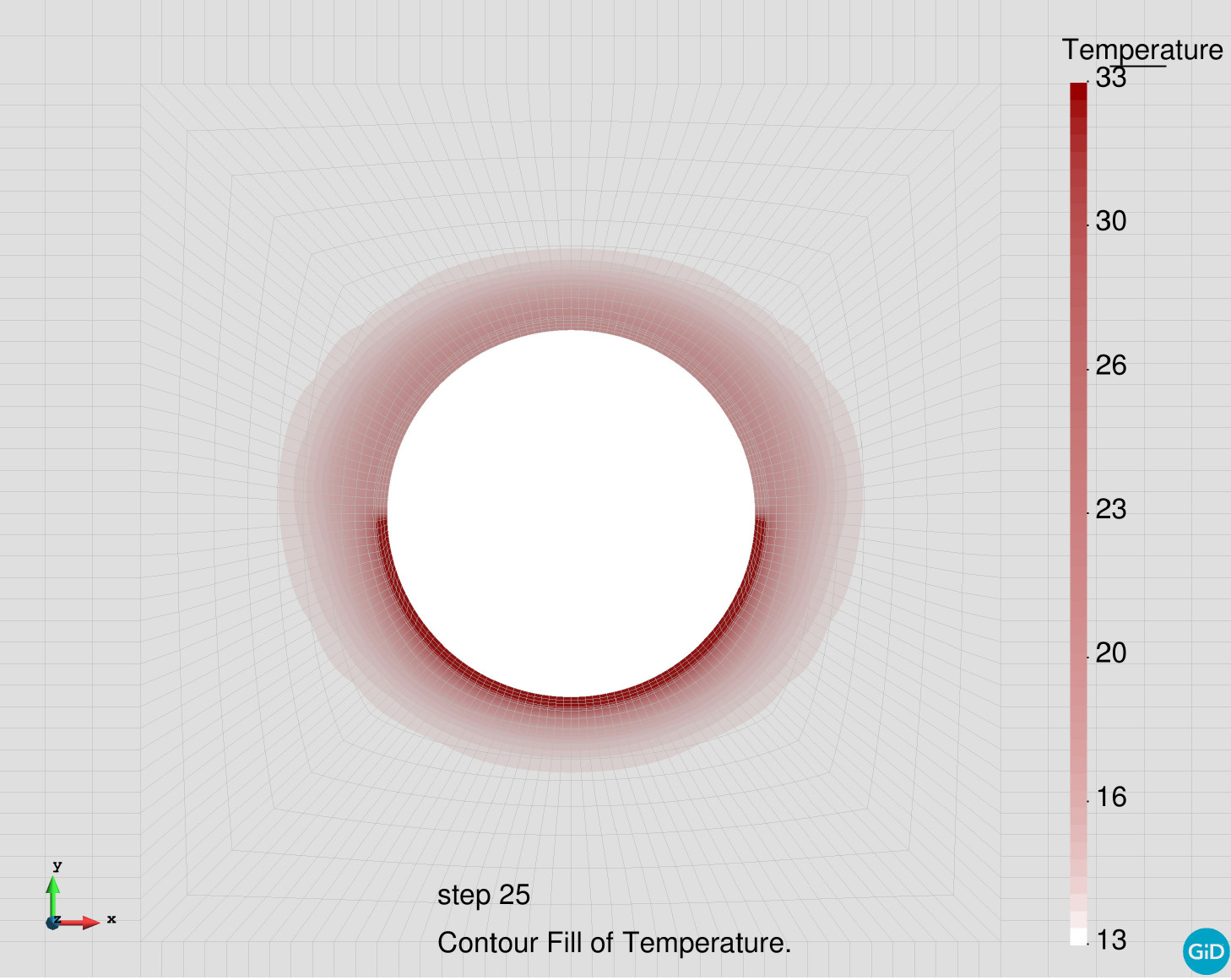} \\
    (a) & (b) \\
    \includegraphics[width=0.45\linewidth,trim=0 58 0 0,clip]{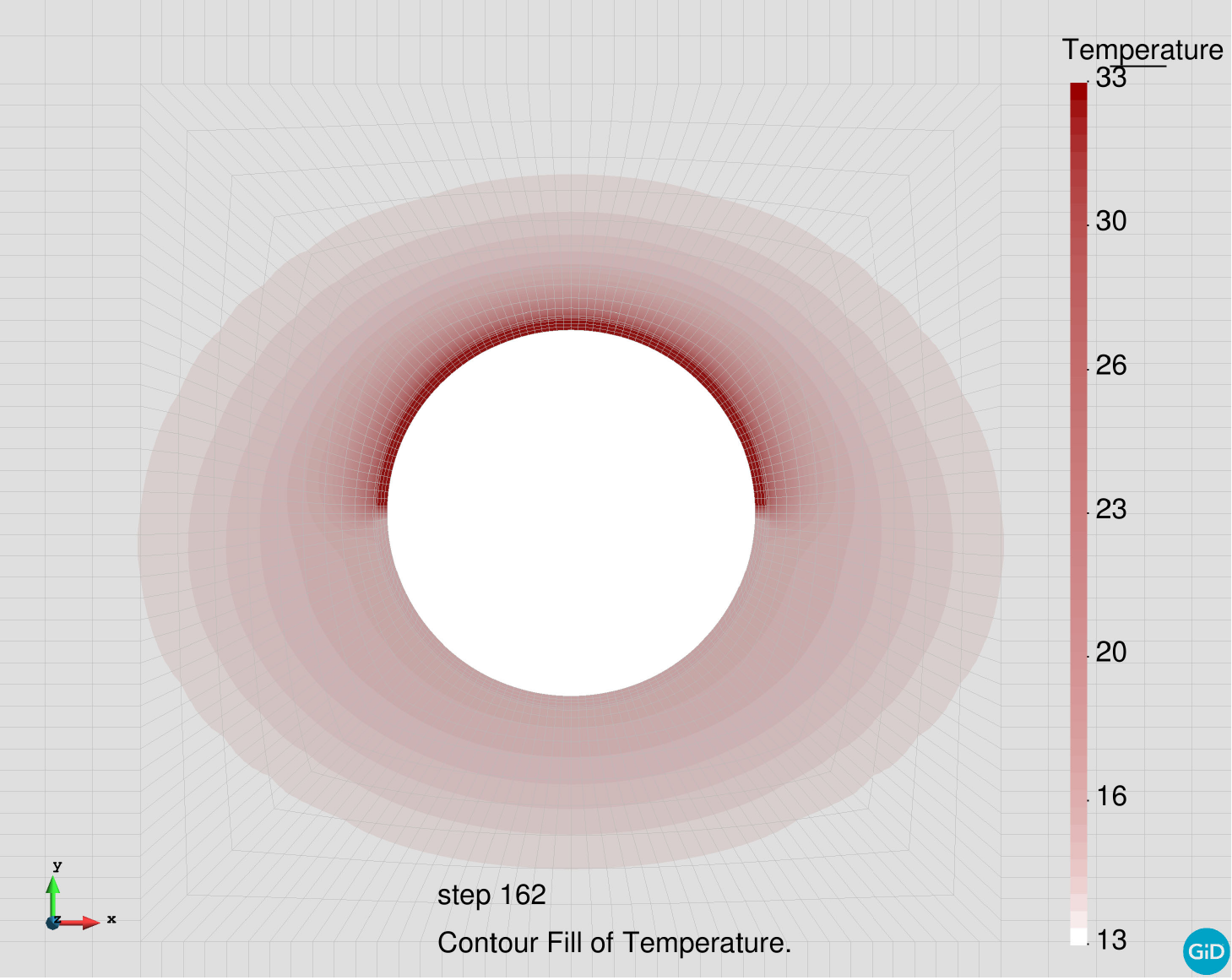} & \includegraphics[width=0.45\linewidth,trim=0 58 0 0,clip]{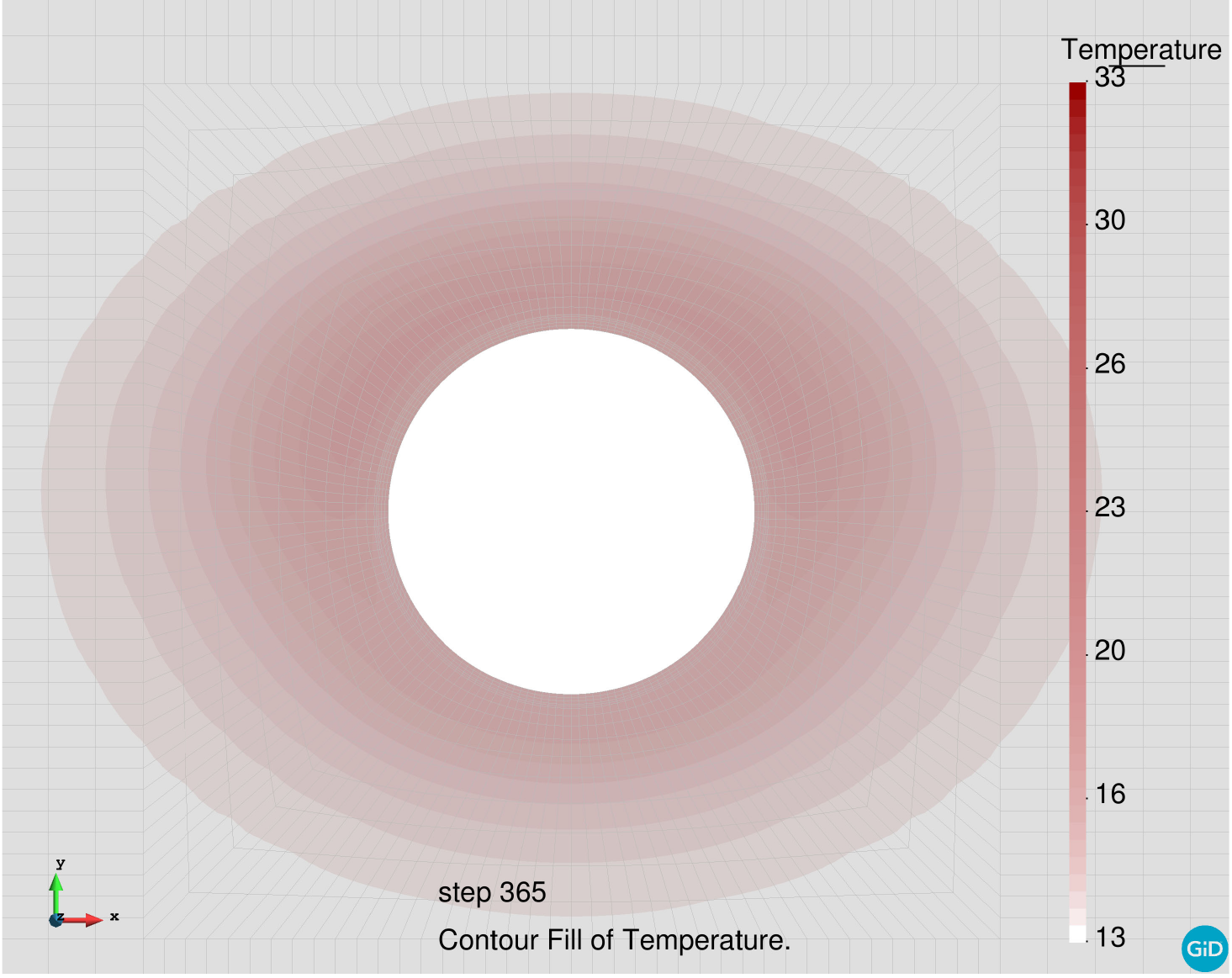} \\
    (c) & (d) 
  \end{tabular}
  \caption{Computed contours of equal temperature in \textdegree C at: (a) 3 days after excavation; (b) 25 days--bottom inner-liner installation; (c) 162 days--crown and sidewall inner-liner installation; and (d) 365 days--end of full tunnel excavation.}
  \label{fig:anisotropic_temperature}
\end{figure}

The anisotropic effects were more noticeable in the temperatures measured in the clay mass in terms of contours at various relevant stages (\cref{fig:anisotropic_temperature}).

\begin{figure}[H]
  \centering
  \includegraphics[scale=1]{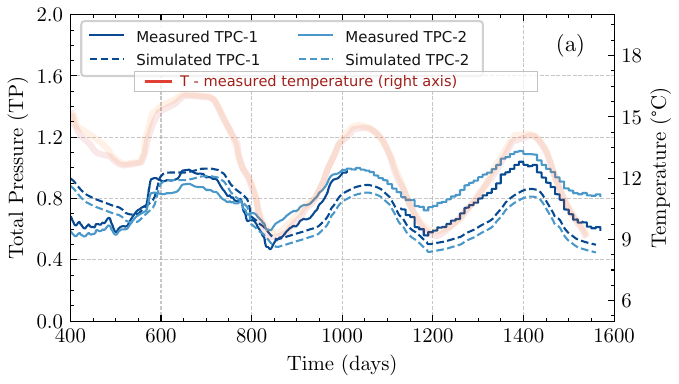}
  \includegraphics[scale=1]{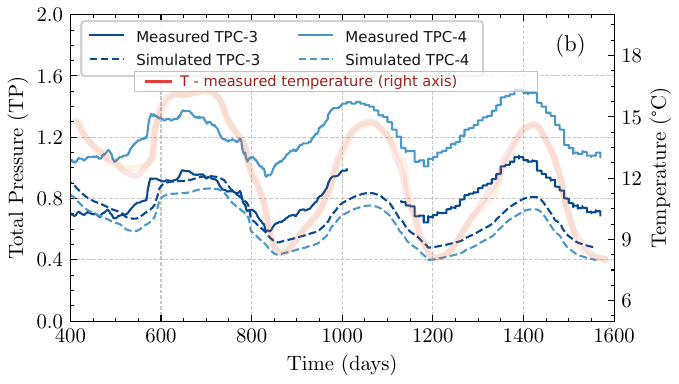}
  \includegraphics[scale=1]{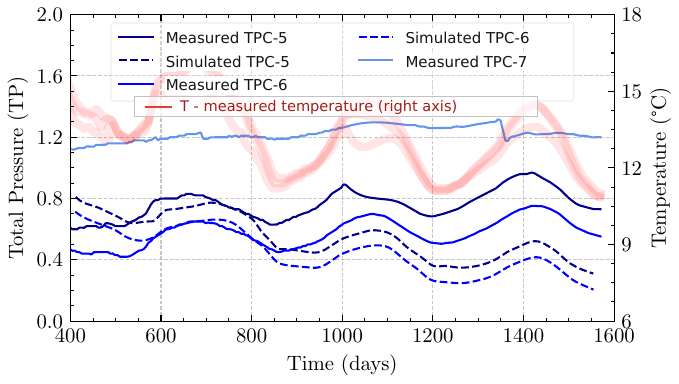}
  \caption{Temperature, $T$, and total radial pressure time series during operation (1~Nov~2017--13~Feb~2021). 
  Panel (a): crown sensors TPC-1 and TPC-2; panel (b): sidewall sensors TPC-3 and TPC-4; panel (c): invert sensors TPC-5, TPC-6, and measured curve of TPC-7. 
  Blue tones denote the pressure (solid: measured; dashed: simulated). 
  The red curves show measured $T$ at the corresponding TPC locations; the right-hand axes report temperature in $^\circ$C. 
  Observed and computed values are compared directly.}
  \label{fig:long_term}
\end{figure}

\Cref{fig:long_term} illustrates the long-term evolution of temperature and total radial pressure after installation of the inner lining. The records exhibited annual temperature cycles, peaking in September and reaching minima in February, consistent with seasonal ambient conditions. The yearly temperature range was approximately 10--14$^\circ$C, with more pronounced fluctuations at the crown and sidewalls and a damped response at the invert. 

These thermal cycles drive the cyclic expansion and contraction of the lining–rock system, resulting in cyclic total pressure variations. Correspondingly, \Cref{fig:long_term} shows pressure oscillations that peak in late summer and reach minima in winter. The typical seasonal sensitivity of pressure to temperature lies between 0.045–0.07~MPa/$^\circ$C for most sensors and is smaller at TPC-7 (0.03–0.045~MPa/$^\circ$C). Superimposed on the seasonal signal, several instruments displayed a slow multi-year pressure increase that was largely independent of temperature, indicating long-term swelling and/or time-dependent deformation of the surrounding rock mass.

The spatial variability has several causes. The model represents the unequal far-field stresses (\(\sigma_v=7\) and \(\sigma_h=5\)~MPa), bedding-controlled stiffness and strength, and the different measured thermal histories applied at the crown, sidewalls, and invert. These factors produce systematic circumferential differences in stress redistribution and thermal loading. The field record also contains more localized effects: re-grouting helps explain the different crown responses at TPC-1 and TPC-2; excavation of CP~5a coincides with the pressure change at TPC-4; and the sustained rise at TPC-7 is consistent with local swelling, grout yielding or cracking, and geological heterogeneity. Segment joints, circumferential variations in grout thickness or properties, local drains, and CP~5a were not represented in the 2D model. The remaining sensor-to-sensor differences therefore cannot be assigned uniquely to any one of these local mechanisms \citep{ZIEGLER-Arash2022}.

The simulations captured the seasonal phasing and the approximate magnitude of the pressure oscillations. They underestimated the long-term pressure growth at several locations. The likely causes are progressive volumetric swelling, local grout and interface damage, and three-dimensional stress redistribution near CP~5a, none of which is resolved by the present 2D formulation. Agreement at locations less affected by these processes supports the use of the model for interpreting the measured thermo-mechanical cycles, but not as a complete prediction of the long-term local response.

Borehole extensometer measurements (\cref{fig:2D_sketch}) provide an independent deformation check on this interpretation. Between December 2016 and April 2021, the reported cumulative radial lengthening was approximately 1.8~mm at the invert array SM-6, 2.8~mm at the eastern array SM-7, and 0.4~mm at the western array SM-8. Deformation at SM-7 was concentrated in faulted intervals and increased after excavation of cross-passage~5a, whereas the invert array continued to deform at roughly 1--2~m depth \citep{ZIEGLER-Arash2022}. The marked directional contrast agrees qualitatively with a heterogeneous, bedding- and structure-controlled response and supports the interpretation that the long-term pressure rise is not purely thermal. A quantitative displacement validation is not claimed because the present section omits cross-passage~5a, discrete faults, and volumetric swelling, and the rock mass displacement data were not used for calibration.

\subsubsection{Discussion on the effect of creep}

\begin{figure}[H]
    \centering
    \includegraphics[scale=1.1]{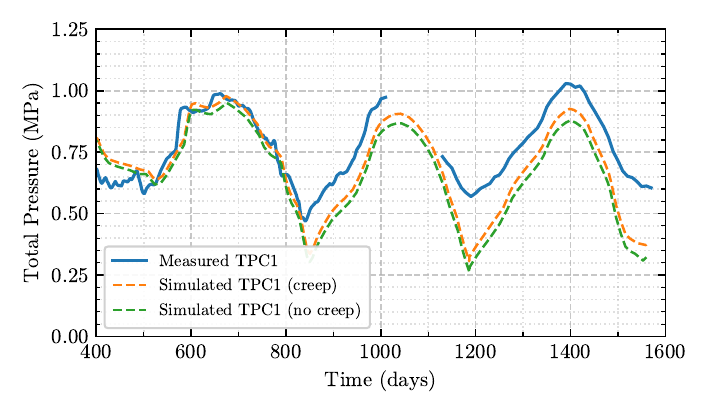}
    \caption{Measured and simulated total radial pressure at TPC-1 versus time. The solid curve shows field measurements; dashed curves show simulations with and without creep.}

    \label{fig:creepeffect}
\end{figure}

The adopted constitutive law includes a viscoplastic (deviatoric) strain-rate term that activates once the shear-stress threshold is exceeded and decays with accumulated viscoplastic strain. This term gradually redistributes the deviatoric stress and therefore yields slightly higher long-term radial pressures than an elastic--plastic formulation. In \cref{fig:creepeffect}, the ``Simulated TPC-1 (creep)'' curve sits above the ``no-creep'' curve, showing that time-dependent shear deformation contributes to delayed stress transfer towards the lining.

Field data show that the total pressure continues to rise after $\sim$800 days while tracking seasonal temperature cycles. The present creep law is purely deviatoric and does not generate time-dependent volumetric strain. Thermal expansion is included, but hydromechanical swelling is not. This omission helps explain why the simulation underestimates the late-time upward drift despite reproducing the seasonal phasing. A moisture- and temperature-dependent volumetric swelling term would be needed to represent the observed monotonic component. Shear creep still contributes to delayed stress redistribution and to sustaining the lining load.

\subsubsection{Scope and limitations of the constitutive representation}

The formulation captures temperature-dependent stiffness and strength, bedding-related anisotropy, deviatoric creep, plastic straining in the EDZ, and a permeability increase tied to the accumulated plastic multiplier. Several processes remain simplified or absent. The creep parameters do not vary explicitly with temperature or saturation, and the creep flow has no volumetric component. The model therefore cannot reproduce anisotropic, moisture-driven swelling. Plastic deformation provides a continuum proxy for EDZ development, but discrete microcracking, fault reactivation, slip along bedding, crack closure or healing, and irreversible changes in crack-network permeability are not resolved. The grout is represented as a homogeneous continuum, and the fully bonded interfaces cannot slip, debond, or open. Finally, the 2D plane-strain geometry omits segment joints, axial variability, cross-passage~5a, and local drainage.

These omissions affect the back-analysis in identifiable ways. Missing swelling and temperature-dependent creep tend to suppress the monotonic long-term pressure increase. Smeared EDZ plasticity and idealized interfaces can redistribute localized loads too uniformly, while omission of CP~5a and drains can shift both the amplitude and timing of individual sensor responses. Calibrated stiffness, strength, creep, or permeability parameters may consequently compensate for processes that they do not physically represent. The parameter set should therefore be interpreted as an effective sectional calibration, with the seasonal thermo-mechanical trends more strongly constrained than the local long-term response.


\subsection{Sensitivity analysis: thermal boundary conditions (post-construction)}
This section examines the period {after} the first year (days $>365$) and tests how alternative temperature histories at TPC-1 affect the {total radial pressure} on the tunnel lining. Let $T_i$ be the measured daily mean at day $i$ and $\bar{T}$ its series mean ($\bar{T}\approx 13~^\circ$C). We define three synthetic histories by scaling the deviation from $\bar{T}$ and adding a small mean shift:
\begin{align}
\text{Scenario 1:}\quad & T^{(1)}_i = \bar{T} + 1.5\,(T_i-\bar{T}), \label{eq:scen1}\\
\text{Scenario 2:}\quad & T^{(2)}_i = \bar{T} + 2.0\,(T_i-\bar{T}), \label{eq:scen2}\\
\text{Scenario 3:}\quad & T^{(3)}_i = \bar{T} + 1.5\,(T_i-\bar{T}) + 1.0. \label{eq:scen3}
\end{align}

The bounds reflect plausible increases in seasonal swings (air exchange/ventilation) and modest mean warming at the portals; they are illustrative rather than predictive. All far-field stresses, hydraulic heads, and material parameters remained identical to the calibrated base case; the initial pore pressure field was hydrostatic and consistent with the boundary heads used in the tunnel-scale model.

\begin{figure}[H] 
\centering \includegraphics[scale=1]{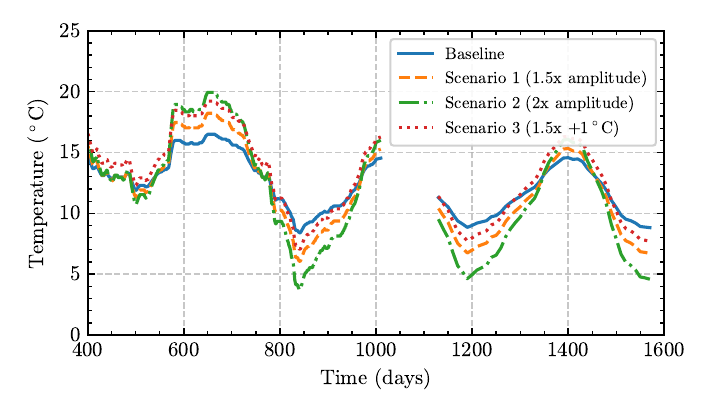}
\caption{Baseline and synthetic TPC-1 temperature histories for the post-construction period (days $>365$).} 
\label{fig:tpc1_temp_scen} 
\end{figure}

After day 365, the baseline varied by approximately $9$–$16~^\circ$C. Scenario~1 produces warmer summers ($\sim\!19~^\circ$C) and cooler winters ($\sim\!7~^\circ$C). Scenario~2 represents an extreme climate (summers $>20~^\circ$C, winters $\sim\!5~^\circ$C). Scenario~3 mirrors Scenario~1 with a $+1~^\circ$C bias (\cref{fig:tpc1_temp_scen}).
 
\begin{figure}[H]
    \centering
    \includegraphics[scale=1]{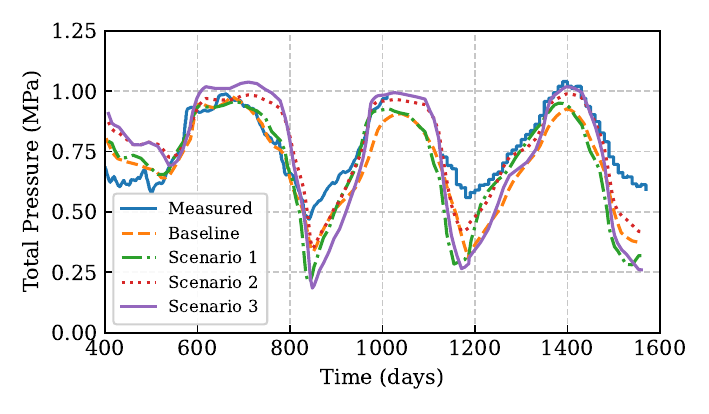}
\caption{Measured and simulated total radial pressure at TPC-1 versus time (days 400–1600).}
    \label{fig:total_pressure_variations}
\end{figure}

~\cref{fig:total_pressure_variations} shows the resulting total pressures at TPC-1 (days 400--1600). Amplifying the thermal swing increases pressure oscillations roughly in proportion to the amplitude, with thermal sensitivities in the range $0.03$--$0.07$ MPa/${}^\circ$C across scenarios. In the fully bonded model, the winter minima deepen more than the summer maxima increase because cooling reduces the compressive thermal strain and radial stress more strongly than heating increases an already compressed state. This result reflects continuum unloading; the model does not simulate physical gap reopening. Scenario~3 shows stronger oscillations and a slight upward baseline due to the $+1~^\circ$C bias. The measured pressure continues to rise although temperature plateaus or decreases, indicating an additional non-thermal contribution such as progressive swelling.

\subsubsection{Pore pressures}

\begin{figure}[H]
  \centering
  \includegraphics[width=1\linewidth]{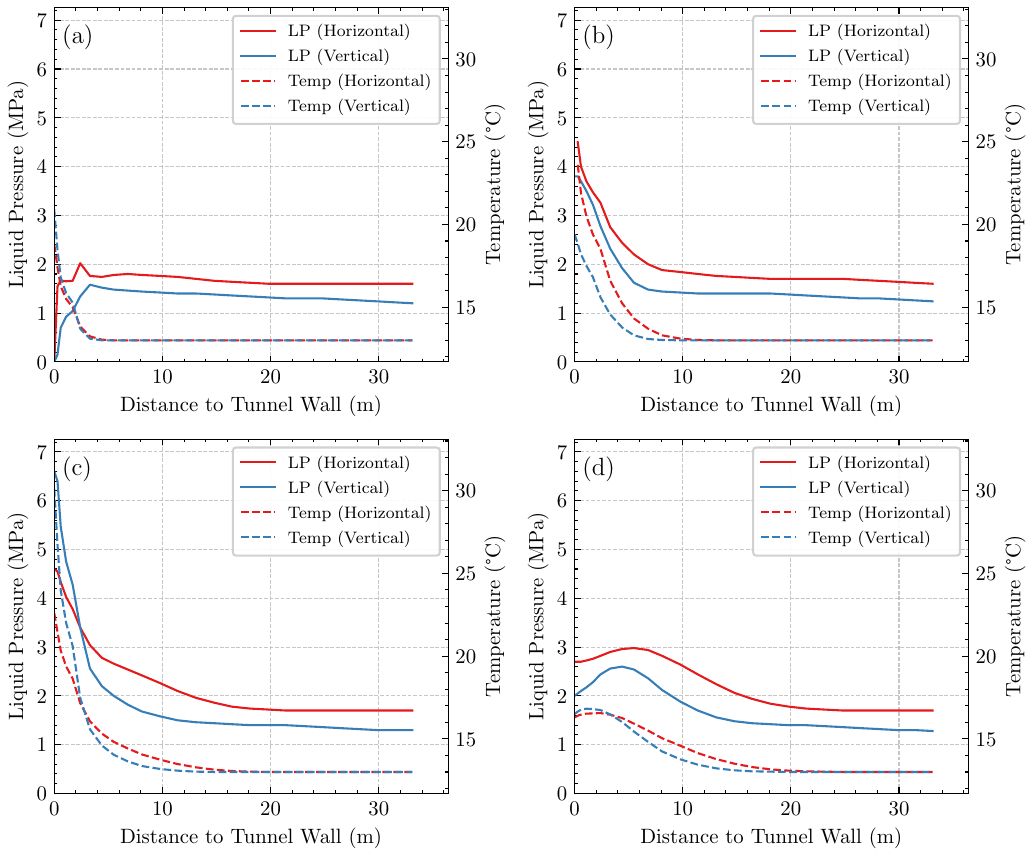}
  \caption{Computed temperature and pore-pressure profiles along two radial transects in
  the 2D cross-section: \emph{transverse} (sidewall; ``horizontal'', $x$) and
  \emph{crown--invert} (``vertical'', $y$). Panels show \text{(a)} day~3,
  \text{(b)} day~25, \text{(c)} day~162, and \text{(d)} day~365.}
  \label{fig:T14}
\end{figure}

\noindent
\Cref{fig:T14} shows the evolution of liquid pressure and temperature along two radial
transects emanating from the tunnel wall: a {transverse} path that intersects the
sidewalls (referred to as ``horizontal'' in the 2D cross-section, $x$–direction) and a
{crown--invert} path (``vertical'', $y$-direction). Immediately after lining installation (day~3), exothermic heat raises pore pressures
near the wall by a few MPa on both transects, with slightly higher and broader peaks
along the transverse path (\cref{fig:T14}a). By day~25, the thermal front has advanced
a few meters, and the near-wall excess pressure remains elevated; the profiles indicate
faster dissipation along the crown--invert path than along the transverse path
(\cref{fig:T14}b). By day~162, the thermal influence extends to $\sim$10–15~m from the
wall in the numerical results (slightly farther along the transverse path), while a pronounced near-wall
overpressure remains on the transverse path (\cref{fig:T14}c). At one year (day~365), the thermal gradient has
{decreased} markedly, yet liquid pressure close to the wall remains higher on the
transverse path (stabilizing around $\sim$3~MPa at the wall and reducing with distance),
consistent with anisotropic dissipation (\cref{fig:T14}d).%

The directional differences in dissipation seen in \cref{fig:T14} highlight the lasting
impact of thermal expansion and anisotropic permeability aligned with the cross-section
geometry: the transverse direction retained higher pressures and relaxed more slowly.
Overall, the relationship between temperature and liquid pressure is most pronounced
near the lining, where heat released during construction and early curing produced the
largest temperature increases and, consequently, the largest excess pore pressures. The modelled
thermal-front extent is not presented as a sensor-validated field measurement.%

The computed multi-megapascal near-field overpressure is strongly conditioned by the
impermeable lining--grout assembly and no-flow rock boundary. In the tunnel, drainage,
hydraulic opening of fractures, and damage-induced permeability changes would provide
additional pressure-release paths and could lower the pressure below the initial value.
Because those processes and longitudinal drainage are absent, the contours should be
interpreted as an idealized sectional response and likely overestimate sustained field
pore pressure near the lining.

\begin{figure}[H]
  \centering
  \includegraphics[width=0.96\linewidth]{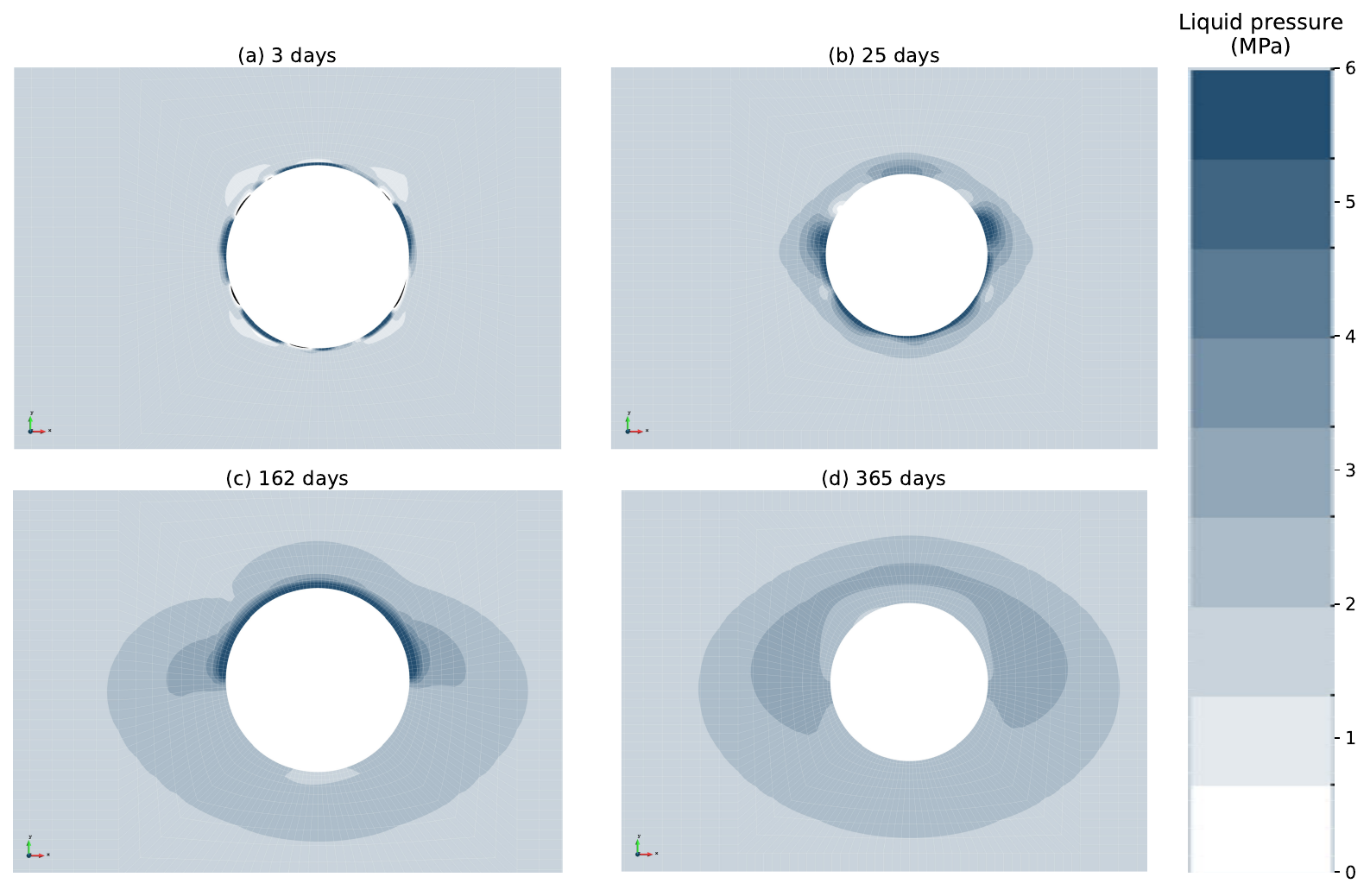}
  \caption{Contours of equal liquid pressure: \text{(a)} day~3,
  \text{(b)} day~25, \text{(c)} day~162, and \text{(d)} day~365. All panels use the shared 0--6\,MPa scale.}
  \label{fig:liquid_pressure}
\end{figure}

\noindent
Contours of equal liquid pressure in \cref{fig:liquid_pressure} further visualize this
anisotropy. The zone of peak excess pressure progressively {shifts outward} from the
wall as heat diffuses, whereas near-wall pressures drop where the rock has cooled. At
365~days, pressures have not yet fully equilibrated in OPA shale: mild under-pressures
persist mainly along the crown--invert transect adjacent to the lining, whereas positive excess pressures remain
at several meters offset, where the temperature changes are more gradual. These results
underline the need to account for coupled temperature–pressure effects in low-permeability,
anisotropic formations, where dissipation is slow and sustained thermal loads can affect
lining actions.%

\subsubsection{Plastic zone}

\begin{figure}[H]
  \centering
  \includegraphics[width=0.52\linewidth]{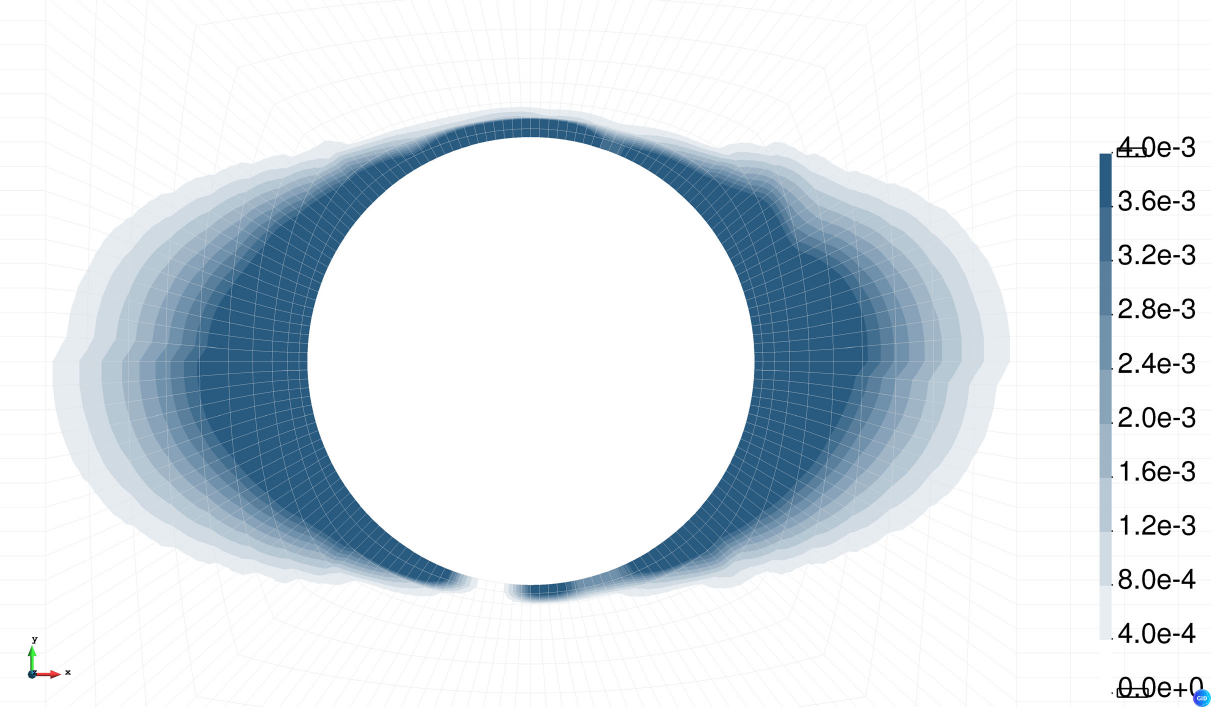}
  \caption{EDZ after $365$\,days in terms of the cumulative plastic multiplier $ \mathrm{EP}_{\mathrm{mult}}$ (dimensionless). The colour bar spans $0$--$4\times10^{-3}$; the maximum displayed is $\approx 4\times10^{-3}$.}
  \label{fig:EDZ}
\end{figure}

\noindent
\Cref{fig:EDZ} maps the cumulative plastic multiplier \(\mathrm{EP}_{\text{mult}}\) after
365 days, a proxy for the intensity of irreversible strain. The rock mass (Opalinus Clay, OPA)
is treated as cross-anisotropic, with stiffness and strength referenced to a material
fabric dipping \(45^\circ\) in the 2D cross-section. This representative model value lies within the variability of field measurements reported near the monitoring section.

As the excavation unloads the cross-section, the principal stresses rotate. The vertical
far-field stress is the maximum principal stress, so excavation produces a strong
compressive stress concentration at the sidewalls. Combined with bedding-dependent
strength, this concentration promotes yielding in the transverse (``horizontal'')
direction and produces lateral lobes of elevated \(\mathrm{EP}_{\text{mult}}\). This pattern
is consistent with conceptual models and observations of layered clay shales near circular
openings, where sidewall lobes are dominated by extensional damage and crown/invert by
slip-type response \citep[e.g.,][]{Martin2002,Bossart2002,thoeny2014,Lisjak2015}.

The formulation captures this directionality through the stress-scaling tensor in \cref{eq:anisotropic_stress},
with \(\sigma_{xx}\) scaled by \(1/C_N\) and \(\sigma_{yy}\) by \(C_N\) (\(C_N=1.33\)). In effect,
the yield threshold is \emph{lowered} at the sidewalls and \emph{raised} at crown and invert,
    driving an asymmetric EDZ with two lateral lobes in the continuum calculation.

\begin{figure}[H]
  \centering
  \includegraphics[width=0.48\textwidth]{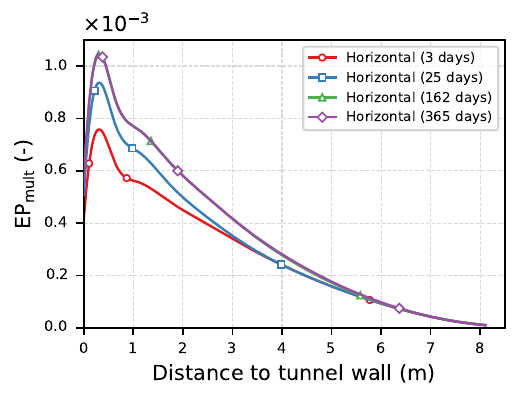}\hfill
  \includegraphics[width=0.48\textwidth]{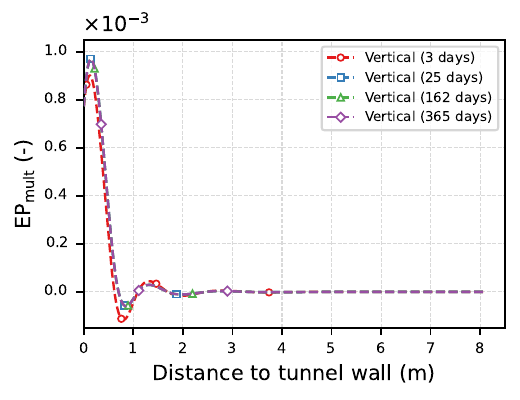}
  \caption{Temporal evolution of $\mathrm{EP}_{\text{mult}}$ profiles by direction. Left:
  transverse (``horizontal'') at 3, 25, 162, and 365~days. Right: crown--invert (``vertical'')
  at the same times. Curves that coincide are visually indistinguishable.}
  \label{fig:EP_H_v}
\end{figure}

\Cref{fig:EP_H_v} presents $\mathrm{EP}_{\text{mult}}$ profiles versus distance from the wall
along the transverse and crown--invert transects at four times. In all cases, plasticity
is concentrated within the near field ($\lesssim$2--3~m) and then decays rapidly. The
transverse profiles display partly higher peaks and a broader influence zone, and they
grow from 3 to 365 days, indicating cumulative redistribution of deviatoric stress in
that plane. The crown--invert profiles remain small and comparatively stable, with only
minor near-wall oscillations that decay within $\sim$1--2 m. Several curves overlap
closely and are therefore visually indistinguishable in the right-hand panel. This
strong difference stems from (i) the anisotropic stress scaling ($C_N$) that disfavors
yielding in the vertical direction and (ii) the rotation of principal stresses that aligns
sidewall stresses with weaker effective orientations. Permeability evolves with plastic straining in the model via the parameter $\eta$
(\cref{tab:model_parameters}), so near-wall plasticity slightly enhances $k$; together
with thermal pressurization this explains the broader, slower-dissipating pressure
response along the transverse path (compare \cref{fig:EP_H_v} and \cref{fig:T14}).

The continuum model does not include discrete faults or bedding-plane slip. Under the
adopted vertical maximum stress and out-of-plane minimum stress, reactivation of these
structures could enlarge the damaged zone at the crown and invert. The contrast between
the SM-6 and SM-8 extensometer records may partly reflect this mechanism, which is not
captured by the calculated plastic multiplier.%

\section{Conclusions}\label{sec_concluding_remarks}

This study presents a thermo-hydro-mechanical analysis of a TBM-excavated tunnel in Opalinus Clay (OPA) using the New Belchen Tunnel as a case study. A coupled 2D plane-strain model was compared with long-term pressure and temperature monitoring to examine the processes controlling the sectional response.

Concrete hydration and seasonal temperature fluctuations explain much of the short- and medium-term pressure variation. The model reproduces the phasing and approximate magnitude of these cycles, but it underestimates the monotonic long-term pressure increase at several sensors. The extensometer records and the pressure-temperature mismatch indicate that swelling, geological structure, and local construction effects also contribute.

Temperature changes generate pore-pressure transients in the low-permeability OPA and alter the effective stress field. Bedding-related anisotropy and unequal far-field stresses produce a directional excavation damage zone and circumferentially variable lining loads. The computed plastic strain and permeability evolution provide a continuum interpretation of these processes, but they do not resolve swelling strains or the behaviour of rock mass fractures such as faults and bedding planes.

For design in argillaceous formations, the results show the value of including the thermal properties of the lining, grout, and rock in a coupled analysis. The same need arises in deep geological repository assessments, although the loading history and boundary conditions differ from those of an infrastructure tunnel.

The main limits are the purely deviatoric, temperature-independent creep law, the absence of volumetric and anisotropic swelling, fully bonded interfaces without grout cracking or debonding, and the omission of segment joints, local drainage, faults, and cross-passage~5a in the 2D geometry. These simplifications can make calibrated continuum parameters compensate for unresolved processes. Future work should use the rock mass displacement data in the calibration, introduce moisture- and temperature-dependent swelling and creep, and test three-dimensional models with explicit construction and drainage features.

\section{Acknowledgments}

The authors acknowledge the financial support of the Luxembourg National Research Fund (FNR) for the EnergWALL project (grant FNR 18978786/ENERGWALL). The computational resources provided by the University of Luxembourg High-Performance Computing Facility are gratefully acknowledged.

We also thank the Swiss Federal Nuclear Safety Inspectorate (ENSI) and the Department of Earth and Planetary Sciences at ETH Zurich for providing access to the dataset used in this study.

\bibliographystyle{apalike}
\renewcommand{\bibname}{References}
\bibliography{tourchi_etal_2024.bib}
\clearpage

\end{document}